\RequirePackage{fix-cm}
\documentclass[natbib]{svjour3}                     %
\smartqed  
\usepackage{graphicx}
\usepackage{epsf}
\usepackage{amssymb}
\usepackage{amsmath}
\usepackage{placeins}
\usepackage{color}
\usepackage{hyperref}
 \journalname{SSRv}
\newcommand{\be}{\begin{equation}}
\newcommand{\ee}{\end{equation}}
\newcommand{\beq}{\begin{eqnarray}}
\newcommand{\eeq}{\end{eqnarray}}
\newcommand\subsun[1]{{$_{\normalsize\odot}$}}

\newcommand{\kms}{~km~s$^{-1}$}

\def\aj{\rmfamily{AJ~}}           
\def\apj{\rmfamily{ApJ~}}         
\def\apjl{\rmfamily{ApJ~}}        
\def\apjs{\rmfamily{ApJS~}}       

\def\aap{\rmfamily{A\&A~}}        

\def\pasa{\rmfamily{PASA~}}       
\def\araa{\rmfamily{ARA\&A~}}     
\def\mnras{\rmfamily{MNRAS~}}     
\def\pasp{\rmfamily{PASP~}}       
\def\apss{\rmfamily{Ap\&SS~}}     
\def\nat{\rmfamily{Nature~}}      

\def\baas{\rmfamily{BAAS~}}       
\let\iaucirc=\iauc

\begin{document}
\title{Peculiar Supernovae}

\titlerunning{Peculiar Supernovae}        

\author{Dan Milisavljevic$^{1}$ \& Raffaella Margutti$^{2}$ }

\institute{
$^1$ Department of Physics and Astronomy, Purdue University, 525 Northwestern Avenue, West Lafayette, IN 47907, USA \\
$^2$ Center for Interdisciplinary Exploration and
  Research in Astrophysics (CIERA) and Department of Physics and
  Astrophysics, Northwestern University, Evanston, IL 60208, USA\\
\email{dmilisav@purdue.edu and rafmargutti@gmail.com}\\
             }

\date{Received: date / Accepted: date}

\maketitle

\abstract

What makes a supernova truly ``peculiar?'' In this chapter we attempt to address this question by tracing the history of the use of ``peculiar'' as a descriptor of non-standard supernovae back to the original binary spectroscopic classification of Type I vs.\ Type II proposed by \citet{Minkowski41}. A handful of noteworthy examples (including SN 2012au, SN 2014C, iPTF14hls, and iPTF15eqv) are highlighted to illustrate a general theme: classes of supernovae that were once thought to be peculiar are later seen as logical branches of standard events.  This is not always the case, however, and we discuss ASASSN-15lh as an example of a transient with an origin that remains contentious. We remark on how late-time observations at all wavelengths (radio-through-X-ray) that probe 1) the kinematic and chemical properties of the supernova ejecta and 2) the progenitor star system's mass loss in the terminal phases preceding the explosion, have often been critical in understanding the nature of seemingly unusual events. 


\section{Introduction}
\label{sec:intro}

Modern transient surveys have uncovered ever-increasing diversity in the observational properties of supernovae (SNe). Well-known surveys include the Palomar Transient Factory (PTF; \citealt{Rau09}), the Panoramic Survey Telescope and Rapid Response System (Pan-STARRS; \citealt{Chambers16}), La Silla-QUEST (LSQ; \citealt{Baltay13}), the Lick Observatory Supernova Search (LOSS; \citealt{Li00}), the Catalina Real-Time Transient Survey (CRTS; \citealt{Djorgovski11}), the All-Sky Automated Survey for Supernovae (ASAS-SN; \citealt{Shappee14}), and the Texas Supernova Search now operating as the ROTSE Supernova Verification Project (RSVP; \citealt{Quimby06}). The original classification system of Type I and Type II proposed by \citet{Minkowski41} has branched considerably from its binary roots and is continually being updated in the face of new objects that bridge subtypes and extend luminosity ranges.

Given the large range of parameter space that a SN progenitor system can occupy, it should not be surprising that a zoo of diverse events will be observed \citep{Heger02,Woosley02,Sukhbold16}. Mass, metallicity, rotation, single vs.\ binary evolution (and if binary, the mass ratio between the companion stars and their initial separation) make for a broad range of possibilities (see, e.g., \citealt{Podsiadlowski92}, \citealt{Yoon10}, \citealt{Claeys11}, and \citealt{Eldridge17}). And yet, ``peculiar'' has a tradition of use in the SN community, and each time non-standard SNe are encountered there is often an element of surprise in the reported analysis. In this chapter we seek to understand why this might be and explore patterns of the use of peculiar as a descriptor of SNe.  

We include in our investigation the many synonyms of peculiar that can be used to describe unexpected phenomena. This includes, but is not limited to, e.g., ``unusual" \citep{Bersten16}, ``unique" \citep{Tominaga05,Maeda07}, or ``Perplexing, Troublesome, and Possibly Misleading'' \citep{Foley10}. Peculiar is typically characterized through non-standard bolometric luminosity, either in peak value (e.g., \citealt{Smith07}) or longevity (e.g., \citealt{Arcavi17}), or it is associated with non-standard spectroscopic emission (e.g., \citealt{Jha06}). These outstanding properties may be attributable to explosion dynamics or, more often, characteristics related to mass loss environment sculpted by the progenitor star system. Peculiar is sometimes used interchangeably with rare. Most certainly there are psychological motivations behind use of the word: use of peculiar (or a synonym) evokes mystery, intrigue, and curiosity more than something that is normal and familiar.  Fundamentally, use of the word reflects {\it incomplete knowledge}.

Here we focus on peculiar supernovae that at the time of discovery did not fit neatly into any single observational classification. We will discuss how time has shown that some events that were once regarded as highly unusual are in hindsight understood as being normal and/or logical extensions of an already well characterized phenomenon.  Our chapter is by no means a complete census of ``extreme'' events and the reader is directed to excellent reviews of core-collapse and thermonuclear supernovae by \citet{Gal-Yam16}, \citet{Parrent14}, and  \citet{Taubenberger17}.

\section{Type Ib -- The ``first" peculiar supernova}
 
The two main classes of supernovae Type I and II (hereafter SNe I and II) were established by \citet{Minkowski41}. Type II have conspicuous hydrogen features and Type I do not. Today we further divide Type I into those that show Si and S lines (Type Ia) and those that do not (Type Ib and Ic).  Narrow emission lines of SNe IIn are associated with interaction with circumstellar environments. 

Generally, SNe Ia are presumed to be from white dwarf progenitor systems and the remaining classes from massive stars (see discussion and references in \citealt{Milisavljevic13-review}). The classifications have remained chiefly spectroscopic, although some information about light curve shape (e.g., IIP and IIL; \citealt{Barbon79}) and bolometric luminosity (e.g.,   superluminous supernovae; \citealt{Gal-Yam12}) have been incorporated. 

It is challenging to accurately pin down the original use of ``peculiar" to describe supernovae in the \citet{Minkowski41} classification scheme. \citet{McLaughlin63} reported observations of SN\,1954A that were conspicuously different from SNe I and II. However, more notable are the observations of SN\,1962I in NGC 1073 by \citet{Bertola64}. The spectral evolution of SN\,1962I was distinctly different from normal Type I, in particular lacking the Si II 6115~\AA\ absorption feature and having an absolute luminosity lower by 1.4 mag compared with normal Type I. 

Two decades later a fusillade of papers published rapidly between 1985-1987 on ``peculiar Type I supernovae'' zeroed in on what we now refer to as Type Ib \citep{WL85,Harkness87,PF87,Wheeler87}. These investigations were spurred on by multiple nearby supernovae (including SN 1984L in NGC 991 and SN\,1983N in M83) that also lacked the 6115~\AA\ feature normally associated with Type I maximum-light spectra and exhibited absolute luminosities down by $\approx 1.5$ mag compared with normal Type I, despite the shape of the optical light curves being similar.  The suggestion that these be called ``Type III'' supernovae was made \citep{Chevalier86}, but the term was not adopted.

\begin{figure}[!htp]
\centering
\includegraphics[width=\linewidth]{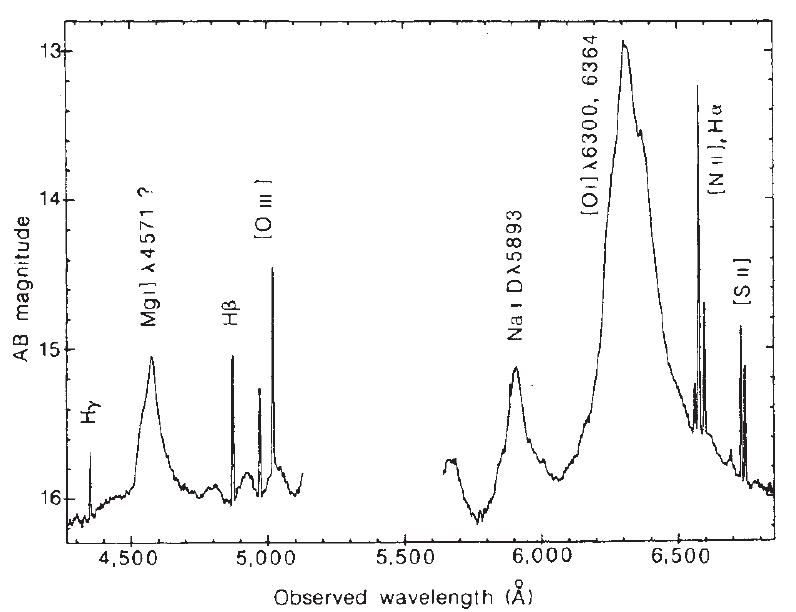}

\caption{Late-time spectrum of SN\,1985F. These ``peculiar'' emissions were in stark contrast with any previously published supernova spectrum. We now know that these broad [O~I], Mg~I], and Na~I lines are standard emissions observed from metal-rich ejecta of stripped-envelope supernovae observed at nebular epochs. Reproduced with permission from \citet{Filippenko85}.}

\label{fig:85F}
\end{figure}

Perhaps most notable of these reports was on SN\,1985F, deemed ``A peculiar supernova in the spiral galaxy NGC4618'' by \citet{Filippenko85}. SN\,1985F exhibited an optical spectrum dominated by very strong, broad emission lines of [O I] 6300, 6364, [Ca II] 7291, 7324, the Ca II near-IR triplet, Mg I] 4571, and Na I D (Figure~\ref{fig:85F}).  The complete absence of hydrogen led to a formal classification of SN I. However, at that time no known spectra of SNe I, which generally exhibit a complex blend of P-Cyg profiles, resembled that of SN 1985F. Many properties of SN\,1985F suggested that the progenitor star was massive and had lost its hydrogen envelope prior to exploding \citep{Filippenko86,Begelman86}, which had been anticipated as a plausible SN progenitor system for the supernova remnant Cassiopeia A a decade previously by \citet{Chevalier76}. 

The question of whether or not SN\,1985F was distinct from SNe Ia and SNe Ib was resolved when \citet{Gaskell86} published a spectrum of the Type Ib SN 1983N obtained eight months past maximum. The two spectra were very similar, and formed a crucial link between SN\,1985F and other peculiar SN I (now known as Type Ib). These data demonstrated that SN 1985F was a SN Ib discovered long after maximum and, more generally, made it clear that SNe Ib transition into a nebular phase dominated by forbidden transitions that is conspicuously different from SNe Ia.  The findings confirmed a previous suggestion by \citet{Chugai86} that SN 1985F might be a SN Ib discovered long after maximum.  The recognition of He I lines in early-time spectra of SNe Ib by \citet{Harkness87}, provided additional evidence that SNe Ib are a physically separate subclass of SNe I initiated by the core collapse of a massive star \citep{WL85}. Wolf-Rayet stars were quickly hypothesized as natural progenitor systems \citep{Begelman86,Gaskell86}.  Late-time observations of Type Ib SN\,1984L \citep{SK89} and detailed analysis of the nebular spectrum of SN 1985F by \citet{Fransson89} further supported this notion. For additional discussion the reader is directed to  \citet{Filippenko97} and \citet{BW17}.

As time passed additional heterogeneity in the Type I family beyond the Ia vs.\ Ib division was recognized. \citet{Filippenko90} reported on a He-poor Type Ib supernova and adopted the nomenclature Type Ic. New Type Ia classifications emerged with SN\,1991T \citep{Filippenko92-91T}, which was more luminous by $\sim 0.5$ mag than regular SNe Ia, and SN\,1991bg \citep{Filippenko92-91bg}, which was less luminous by $R \sim 1$ mag compared to normal SNe Ia at peak. Diversity continues to be recognized to this day; e.g., the class is  SN 2002cx-like Type Ia supernovae (also known as SNe Iax) represent one of the most numerous peculiar SN classes \citep{Foley13}. SNe Iax differ from normal SNe Ia by having a wide range of fainter peak magnitudes, faster decline rates and lower photospheric velocities. We refer the reader to \citet{Parrent14} for overview of SNe Ia spectra, especially for evolution into the nebular phase. \citet{Taubenberger17} is another excellent source about extremes of SNe Ia.

\begin{figure}[!htp]
\centering
\includegraphics[width=\linewidth]{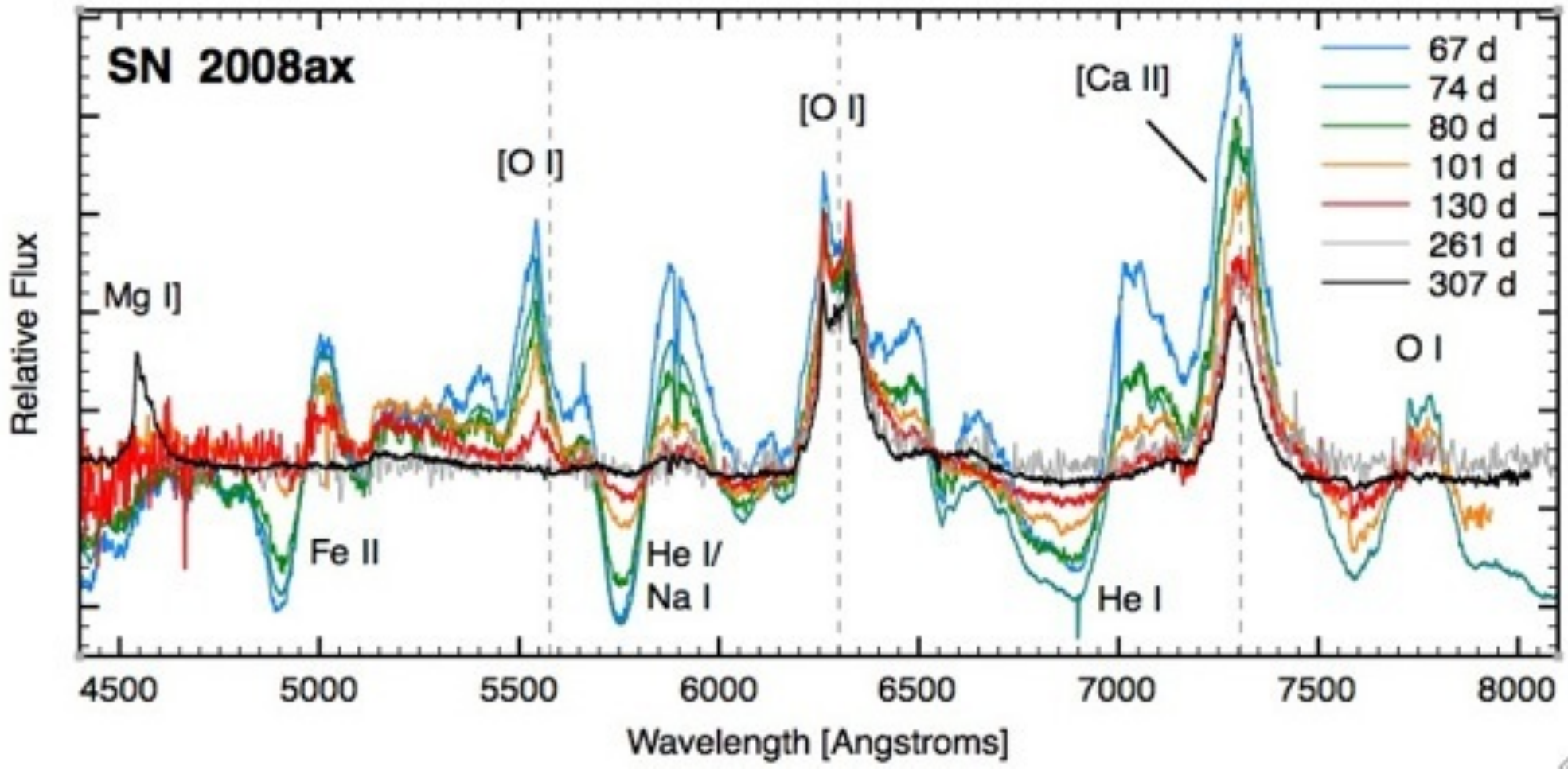}

\caption{Representative evolution of a stripped-envelope supernova from photospheric to nebular epochs. As the supernova expands the P-Cyg profiles originating from the thin expanding photosphere subside to broad emission from forbidden lines originating from the metal-rich ejecta being heated by $^{56}$Co decay. Data are for the Type IIb SN\,2008ax from \citet{Milisavljevic10}.}

\label{fig:08ax}
\end{figure}

The evolving interpretation of the \citet{Filippenko85} observations of SN\,1985F is illustrative. In hindsight this event was not so peculiar after all and was no more than spectra of an otherwise normal SN Ib observed several months after explosion. It was, however, perceived as peculiar because observations broke into an unexplored phase space (several months after explosion in this case). In the decades that have followed, larger aperture telescopes along with improvements to detector and instrument efficiencies have made it possible to routinely follow supernovae from explosion to nebular epochs (Figure~\ref{fig:08ax}). Dozens of high-quality spectra of supernovae at epochs $t > 200$ days post explosion are now available \citep{Matheson01,Modjaz08,Maeda08,Taubenberger09,Black16,Graham17}, and in some cases (e.g., SNe 1957D, 1970G, 1979C, 1980K, 1993J) supernovae have even been monitored several years to decades after explosion at optical wavelengths \citep{FB88,Long89,Turatto89,Milisavljevic12}. These observations show that spectra of SNe Ia and SNe Ib/c evolve in very different manners: SNe Ia consist of many forbidden transitions of Fe and Co, whereas SNe Ib/Ic are dominated by O, Mg, Ca, a mix of Na and He, and other intermediate mass elements. The divergence is a valuable method of spectroscopic fingerprinting supernova progenitor systems in cases when photospheric spectra dominated by P-Cyg absorptions provide ambiguous identifications (see Section \ref{sec:examples}). 

\section{Late-time observations: optical, radio, and X-ray}

Data outside of optical wavelengths have played crucial roles in constraining the nature of peculiar objects. Ultraviolet, optical, and near-infrared wavelengths trace the bulk of the stellar ejecta, while radio and X-ray wavelengths trace emission from the SN's interaction with the local circumstellar environment. Considerable insight into the mass loss environment of the original peculiar Type I (i.e., Type Ib) objects came from observations at radio frequencies; e.g. SN\,1983N \citep{Sramek84} and SN\,1984L \citep{Panagia86}.   Asymmetries in the progenitor star's mass loss can be a major factor contributing to its multi-wavelength emissions and overall degree of ``peculiarity'' (e.g, \citealt{Anderson17,Bilinski17}). Notably, many of the aforementioned evolved SNe still detectable at optical wavelengths (e.g., SN\,1993J) are also associated with long-lived X-ray and radio emission from interaction between the SN and circumstellar material (CSM) \citep{Weiler07}. We discuss additional examples of such SN-CSM interaction in Section \ref{sec:14C}.

Thermal X-ray emission is one of the clearest indications of circumstellar interaction in supernovae \citep{CF16}. The intensity of the emission depends on the square of the density, and thus is a good estimator of the density of the ambient medium, as long as the emission is not absorbed by the medium. So far, over 60 SNe have been detected in X-rays. All of them until recently had been core-collapse SNe, where the CSM is formed by mass loss from the progenitor star. Deep searches have been conducted for Type Ia explosions with null detections and strong constraints both from radio and X-ray observations \citep{Margutti12,Margutti14J,Chomiuk16}, suggesting that these explosions occur in ``clean'' environments ($n < 3$ cm$^{-3}$ at $R \sim 10^{16}$\,cm).

Recently, there has been a successful detection of a special case of Type Ia explosion understood to be interacting with H-rich environments. These are known as SN 2002ic-like or Type Ia-CSM events \citep{Wang04,Silverman13}. Late-time (500-800 days after discovery) X-ray detections of SN 2012ca in {\it Chandra} observations were reported by \citet{Bochenek18}, who favor an asymmetric medium with a high-density component ($n > 10^{8}$ cm$^{-3}$). This finding provides critical insight into the debate as to whether all SNe Ia-CSM are thermonuclear explosions \citep{Fox15,Inserra16}.

\subsection{Supernova metamorphosis: {\it SN\,2014C}}
\label{sec:14C}

\begin{figure}[!htp]
\centering
\includegraphics[width=\linewidth]{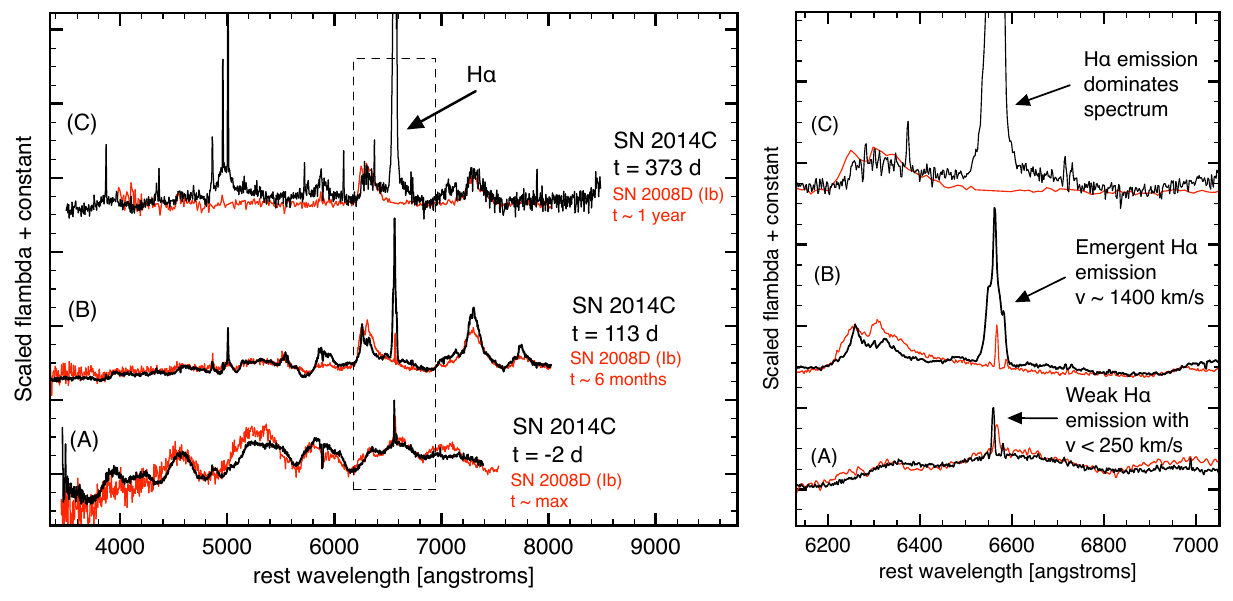}

\caption{Real-time monitoring of the delayed interaction between an H-poor star's supernova explosion (SN 2014C) and its previously stripped H-rich envelope. (A) A spectrum of SN\,2014C obtained around the time of maximum light compared to the H-poor type Ib SN\,2008D. (B) Months later, SN 2014C exhibits typical type Ib emissions originating from O- and Ca-rich inner ejecta, but also unexpected new and extended (FWHM $\approx$ 1400 km/s) H$\alpha$ emission normally only seen in strongly interacting SNe. (C) One year after explosion, H$\alpha$ emission dominates the spectrum. Also seen are low and high ionization emissions of narrow to broad widths originating from many regions (see Fig.\ 3 for enlargement). SN 2014C represents the first time a type Ib SN has been seen to slowly evolve into a strongly interacting Type IIn. Adapted from \citet{Milisavljevic15}.}
\label{fig:14C}
\end{figure}

The remarkable SN 2014C is a closely studied, remarkable example of how late-time SN-CSM interaction can probe fundamental aspects of supernova progenitor systems. SN 2014C was discovered in the nearby spiral galaxy NGC 7331 (D $\sim 15.8$ Mpc) on 5 January 2014 and identified as a comparatively normal Type Ib. However, when observations resumed in May 2014 after SN 2014C had emerged from behind the Sun, its multi-wavelength properties had become notably different from those of normal SNe Ib. Optical spectra showed that conspicuous emission centered around the H$\alpha$ line with an overall full-width-half-maximum (FWHM) velocity dispersion of approximately 1400 km s$^{-1}$ had emerged (\citealt{Milisavljevic15}; see Figure~\ref{fig:14C}). This is a feature associated with radiative shocks in dense clumps of CSM normally seen only in Type IIn SNe.  

Subsequent observations have demonstrated that this metamorphosis is consistent with a {\it delayed} interaction between an H-poor star's supernova explosion (SN\,2014C) and a massive H-rich envelope that had been stripped decades to centuries before core collapse. Coordinated observations with {\it Swift}, {\it Chandra}, {\it NuSTAR}, the {\it Hubble Space Telescope}, and the Karl G.\ Jansky Very Large Array (VLA) captured the evolution in detail and revealed the presence of a massive shell of $\sim 1 M_{\odot}$ of hydrogen-rich material at $\sim6 \times 10^{16}\,\rm cm$ from the explosion site (\citealt{Margutti17-14C}; see Figure 4). The IR luminosity of SN 2014C stayed constant over 800 days, possibly due to strong CSM interaction with the H-rich shell, which is rare among stripped-envelope SNe, and suggests that the CSM shell originated from an LBV-like eruption roughly 100 years pre-explosion \citep{Tinyanont16}. The radio light curve exhibited a double bump morphology indicating two distinct phases of mass-loss from the progenitor star with the transition between density regimes occurring at 100-200 d \citep{Anderson17}. This reinforces the interpretation that SN 2014C exploded in a low-density region before encountering a dense hydrogen-rich shell of circumstellar material that was likely ejected by the progenitor prior to the explosion.  

\begin{figure}[!htp]
\centering
\includegraphics[scale=0.5]{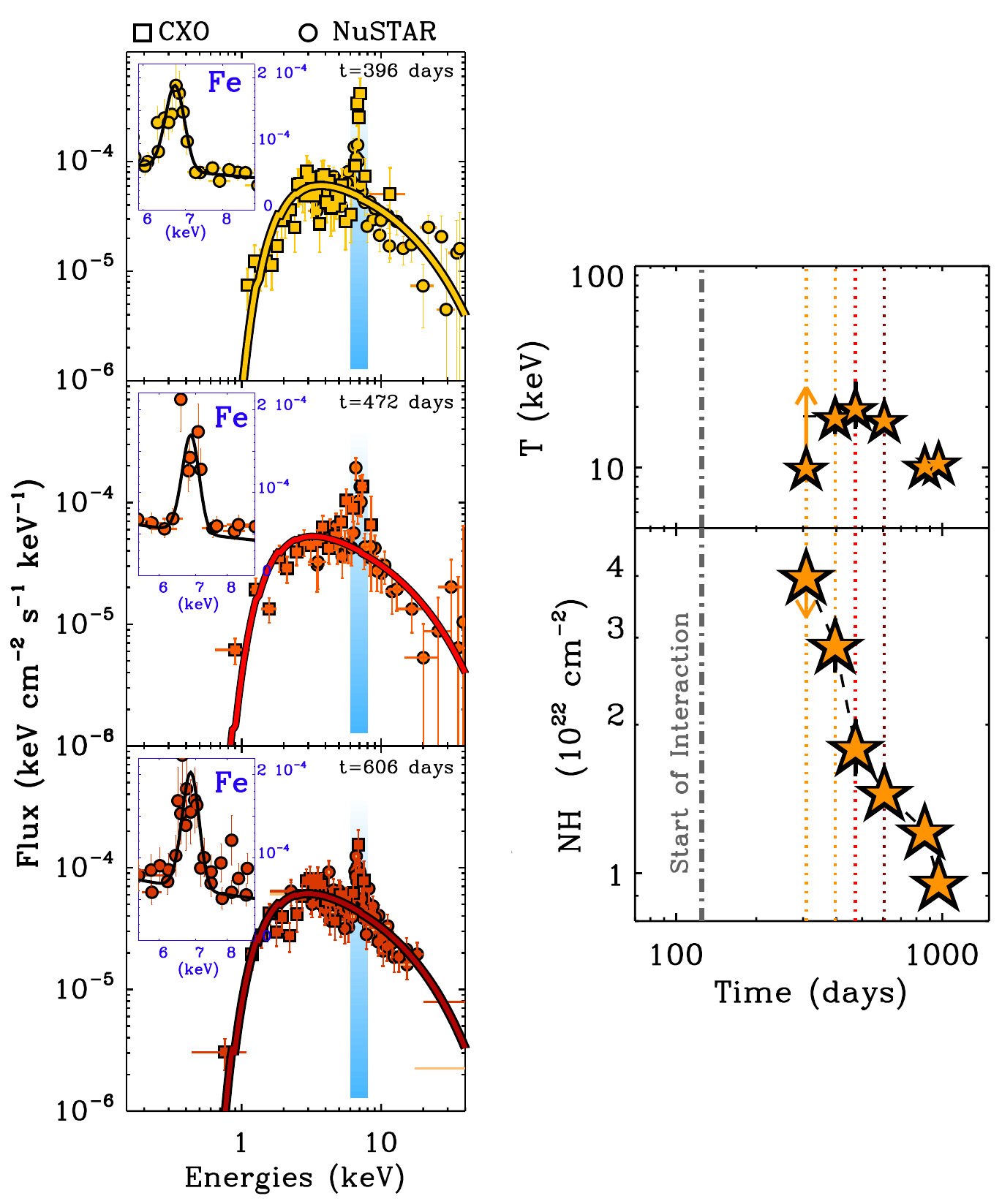}
\caption{SN2014C is the first young extragalactic SN for which soft X-rays ({\it Chandra} - CXO) and hard X-rays (with {\it NuSTAR}) emission were detected and monitored. These observations reveal a thermal spectrum (left panels) with evolving temperature (T, right upper panel) and absorption (NH, lower right panel). Updated from \citet{Margutti17-14C}, with the most recent {\it Chandra}-{\it NuSTAR} observations obtained $\sim1000$ days since explosion.}
\label{fig:14CNustar}
\end{figure}

SN\,2014C is a variant of Type IIn and Ibn SNe that exhibit strong shock interaction between their ejecta and pre-existing, slower-moving CSM \citep{Foley07,Pastorello07,Smith16}, and part of larger family of supernovae from H-poor progenitor stars that interact with H-rich CSM months to years after explosion. This includes SN\,1986J \citep{Rupen87,Milisavljevic08}, the ``Wild cousin of SN\,1987A'' SN\,1996cr  \citep{Bauer08}, and SN\,2001em \citep{Chugai06}, which was at first speculated to be a $\gamma$-ray burst seen off-axis \citep{Soderberg04,Granot04}. Observations of the majority of these supernovae bridging classifications largely missed the transition from one SN type to another. SN\,2014C was unique in the level of details achievable in multi-wavelength follow up across the electromagnetic spectrum.

\begin{figure}[!htp]
\centering
\includegraphics[width=\linewidth]{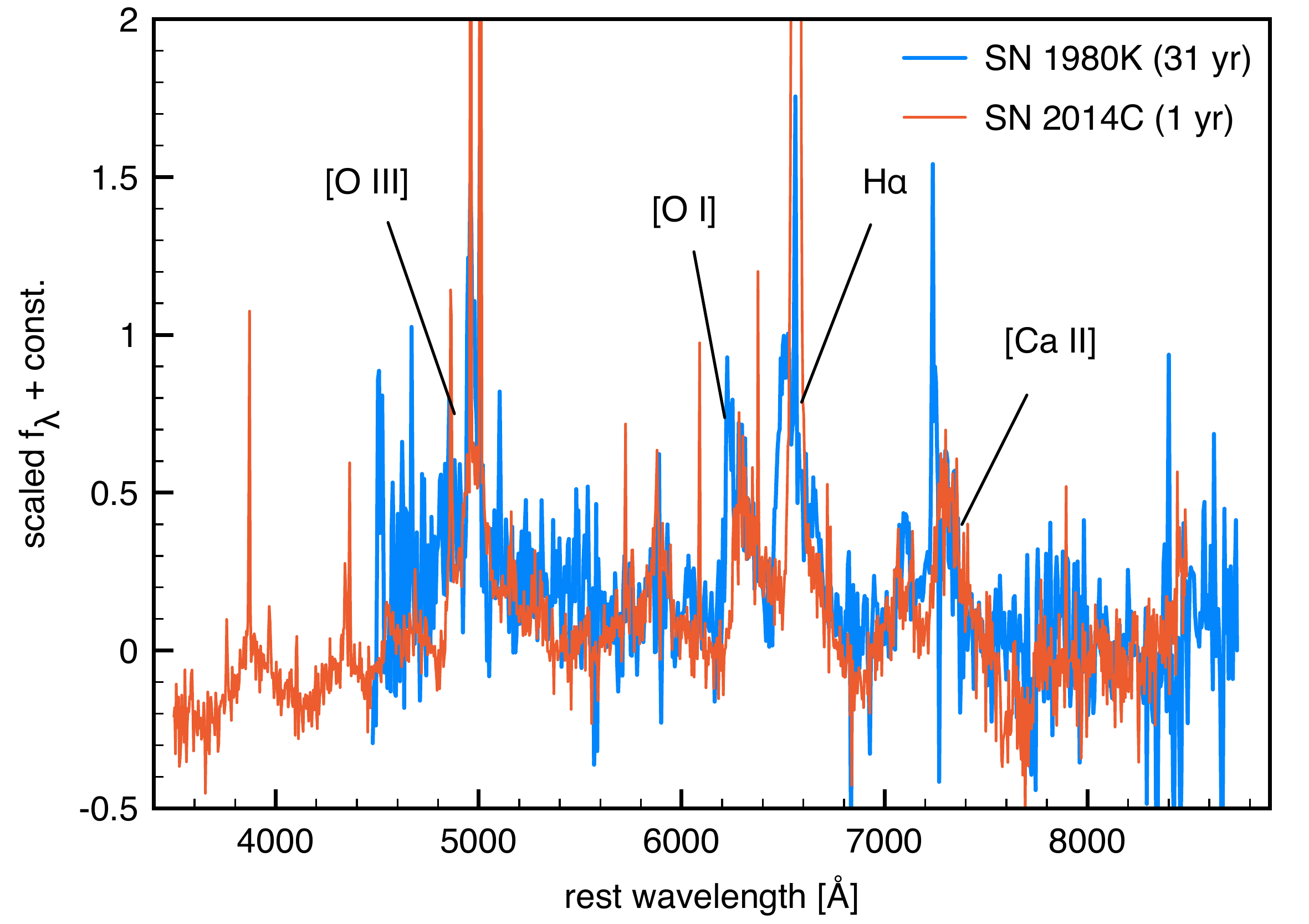}

\caption{Optical spectra of SN\,2014C \citep{Milisavljevic15} compared to SN\,1980K \citep{Milisavljevic12}. Interaction with a dense H-rich CSM shell accelerated the SN-to-SNR evolution (c.f.\ \citealt{MF17}) of SN\,2014C such that at one year its reverse shock had developed strong enough to excite metal-rich material. Shown here is a comparison to the Type II SN\,1980K at 31 yr.}

\label{fig:14C-80K}
\end{figure}

The metamorphosis emphasizes how SN-CSM interaction can dramatically change SN emission features. Despite originally being an H-poor Type Ib, after only one year SN\,2014C rapidly evolved to resemble the H-rich Type II SN\,1980K (Figure~\ref{fig:14C-80K}) at three decades of age.  The overall phenomenon poses significant challenges to current theories of massive star evolution, as it requires a physical mechanism responsible for the ejection of the deepest hydrogen layer of H-poor SN progenitors to be synchronized with the onset of stellar collapse. Although the brief timescales between eruption and supernova explosion have been anticipated in special cases of very massive stars \citep{QS12}, a growing number of systems are being discovered that fall outside most theoretical regimes  \citep{Margutti14,SmithArnett14}. Binary interactions and/or instabilities during the last stages of nuclear burning extending all the way to the core C-burning phase in massive stars are believed to be potential triggers of precursor mass loss (\citealt{Meakin06,Meakin07,Arnett11a,Arnett11b}; see also discussion in \citealt{Margutti17-14C}).

\section{Peculiar supernovae of debated origin}
\label{sec:examples}

\subsection{The SN Ib,c--GRB--SLSN Connection: {\it SN\,2012au}} 
\label{sec:12au}

Twentieth century samples of well-studied SNe were gathered primarily through surveys that targeted known, bright, nearby galaxies. Most of the SNe discovered this way had absolute peak magnitudes fainter than about $\approx -20$ (see e.g., results in \citealt{Jha06,Richardson06}). The latest generation of searches, however, have begun gathering SNe via untargeted surveys that are not biased in this way. One surprising result is that a significant percentage of SNe found in dwarf galaxies peak at $-21$ magnitude or brighter. These superluminous supernovae (SLSNe) are a factor of $10-100$ times brighter than normal core-collapse supernovae. The energy emitted in optical light alone rivals the total explosion energy available to typical core-collapse supernovae ($> 10^{51}$\,erg).  SLSNe are broadly differentiated as hydrogen-rich SLSN-II (e.g., \citealt{Smith07}), and hydrogen-poor SLSN-I (e.g., \citealt{Quimby11}). A further classification for radioactively powered events SLSN-R has also been suggested \citep{Gal-Yam12}.

Though some SLSNe may be powered mainly by the radioactive decay of $^{56}$Ni, many reach peak luminosities far too great to be reasonably explained this way. Interactions between SN ejecta with pre-SN winds is a natural explanation for some systems \citep{Smith07}, but others  show no obvious signs of such ongoing interactions or are unclear \citep{Inserra13,Nicholl13,Lunnan16}. Fundamentally different engines may be powering these observationally distinct events. A leading candidate is energy injection  from a rapidly rotating neutron star (i.e., magnetar) that gets transferred to the kinetic energy of the SN \citep{Kasen10,Woosley10}. 

\begin{figure}[!htp]
\centering
\includegraphics[width=0.49\linewidth]{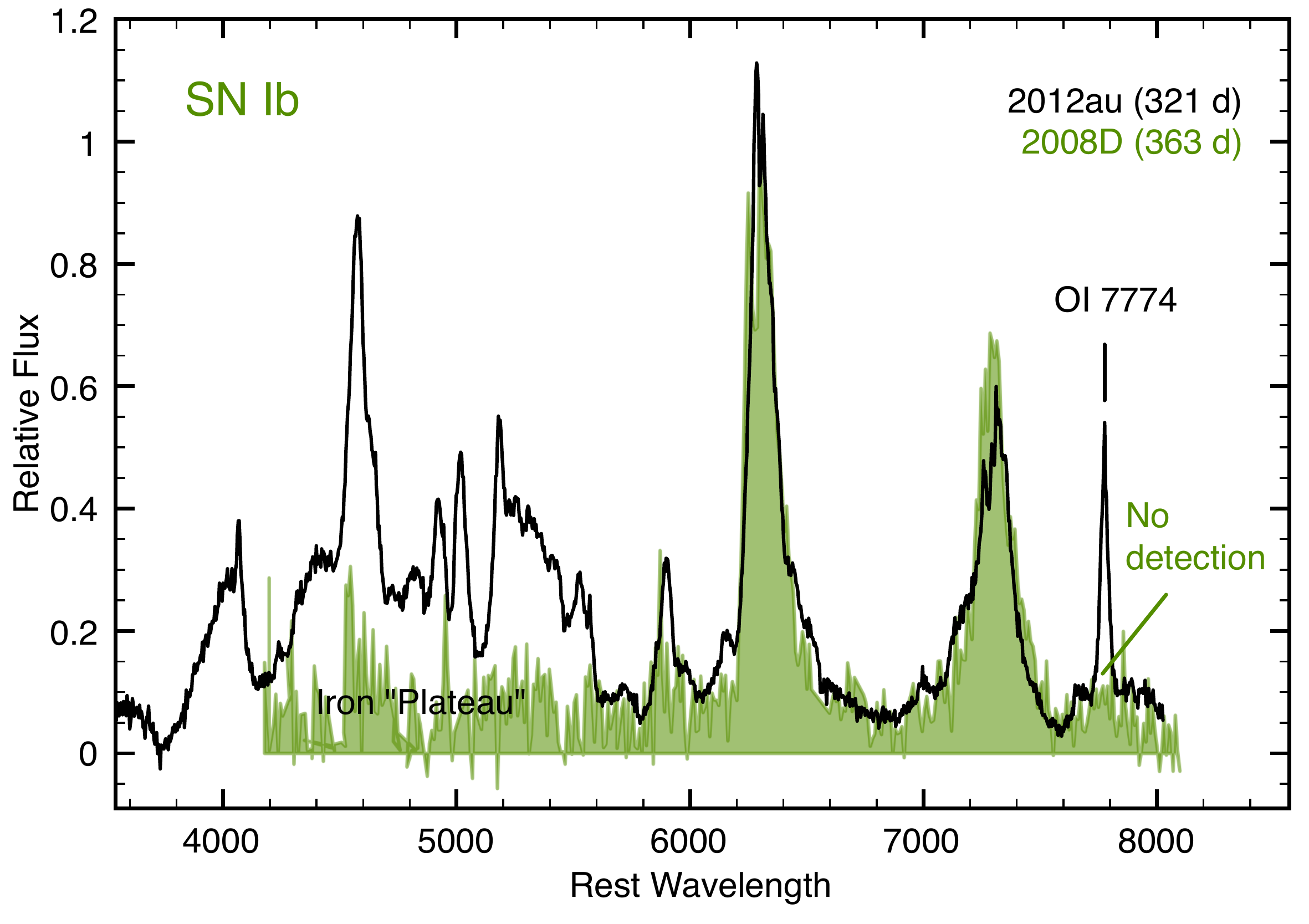}
\includegraphics[width=0.49\linewidth]{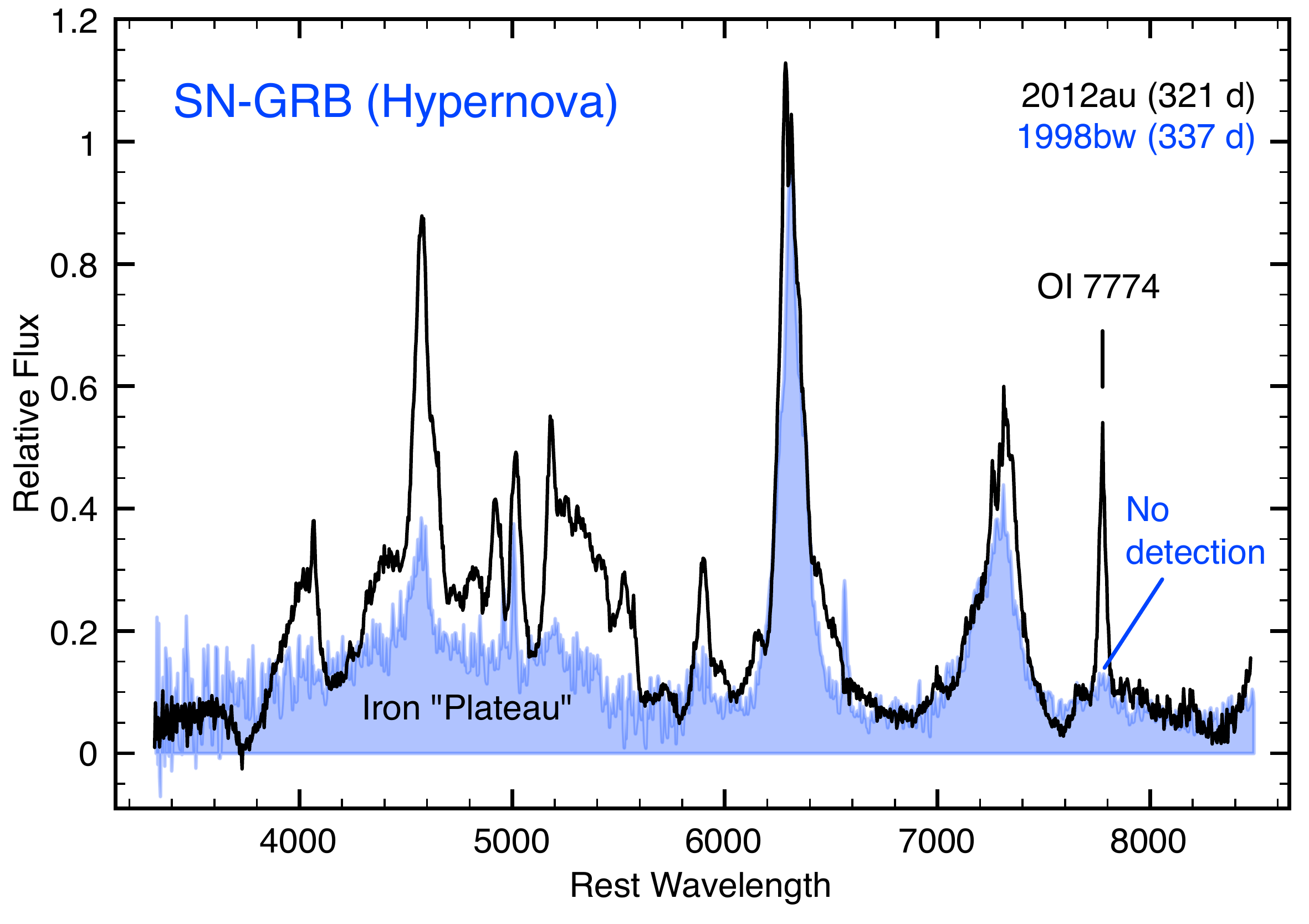}\\
\includegraphics[width=0.49\linewidth]{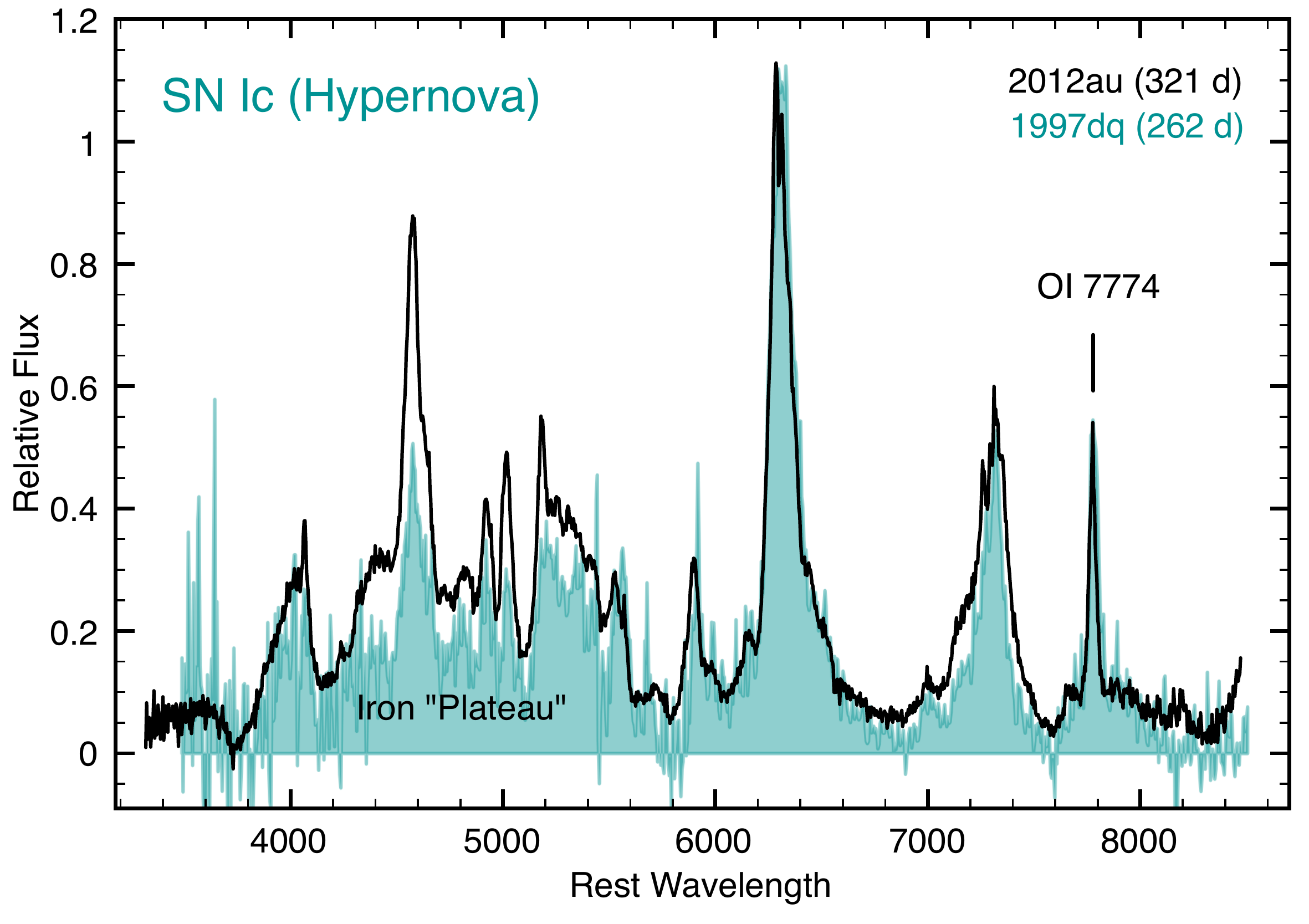}
\includegraphics[width=0.49\linewidth]{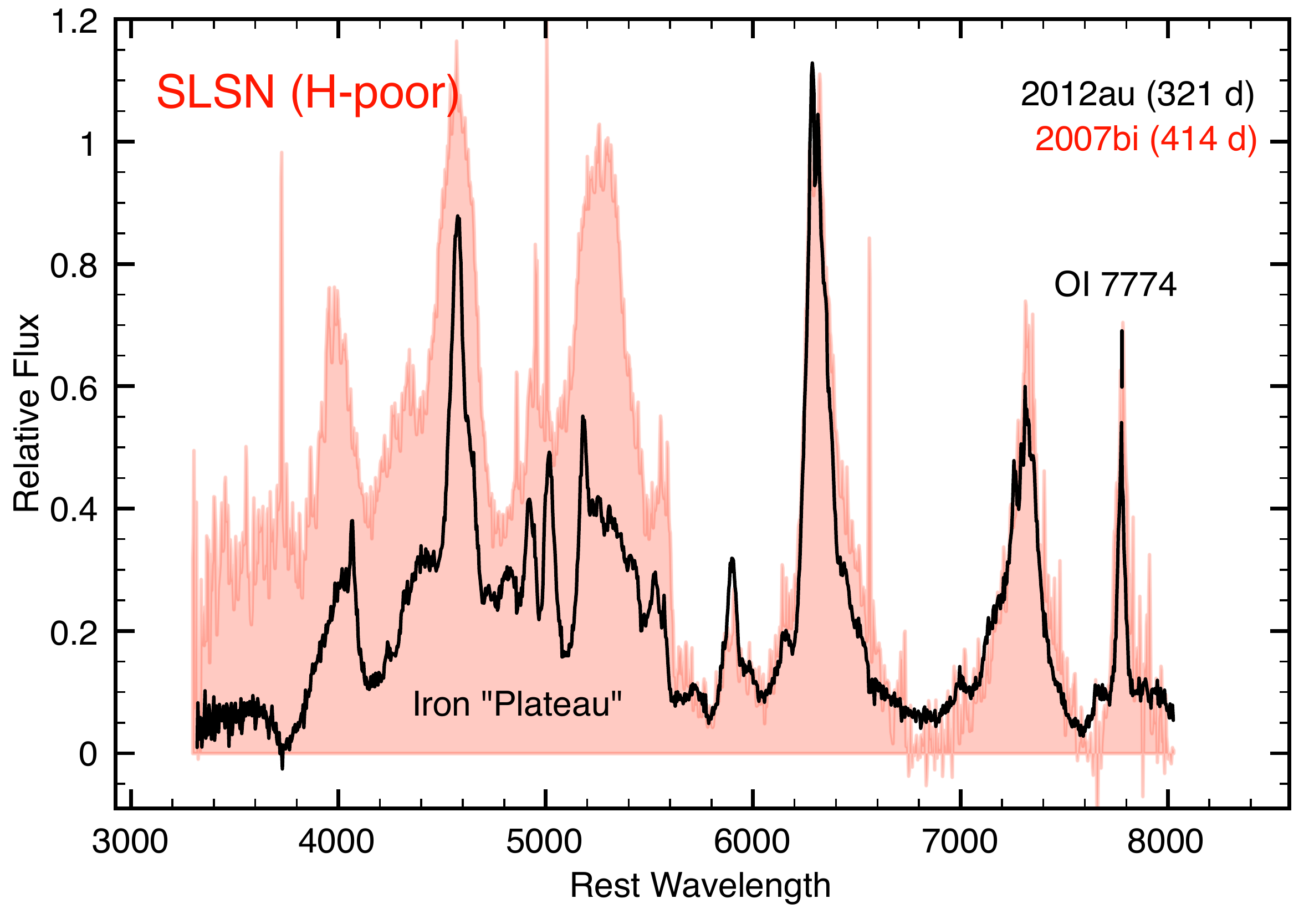}

\caption{Spectral fingerprinting SN 2012au. The top panels show the SN Ib SN 2008D \citep{Tanaka09} and the hypernova SN 1998bw \citep{Patat01}, where spectroscopic evolution is normal and O~I $\lambda$7774 is not detected. These SNe are representative of late-time emissions from the majority of SNe Ib/c. The bottom panels show the hypernova SN 1997dq \citep{Matheson01} and the SLSN
SN\,2007bi \citep{Gal-Yam09}, where spectroscopic evolution is slow, strong O~I $\lambda$7774 emission is detected, and emission from Fe lines forms an emission plateau between 4000 and 5600 \AA. This connection between SNe Ib/c and SLSNe initially made in \citet{Milisavljevic13-12au} was strengthened with late-time observations of the SLSN SN\,2015bn \citep{Nicholl16}. Adapted from \citet{Milisavljevic13-12au}.}

\label{fig:sn2012au}
\end{figure}

Numerous connections have been established between SLSNe and other classifications. The similarity between the spectra of SN 2010gx (SLSN-I) and those of broad-lined Type Ic SNe first hinted that the two classes potentially had similar progenitor star systems  \citep{Pastorello10}. This connection was further strengthened and extended to lower-luminosity counterparts with discovery of SN 2012au, an energetic explosion having a rarely observed combination of late-time properties linking subsets of energetic and H-poor SNe with SLSNe. Initial spectroscopic observations of SN\,2012au showed prominent helium absorption features of an otherwise ordinary Type Ib supernovae. However, continued monitoring through to nebular stages ($t > 250$ days) revealed extraordinary emission properties in its optical and near-infrared spectra not unlike those observed in SLSNe \citep{Milisavljevic13-12au}. 

SN\,2012au had a large explosion kinetic energy of $\sim 10^{52}$ erg and $^{56}$Ni mass of $\approx 0.3 M_{\odot}$ on par with SN\,1998bw \citep{Takaki13}. Its intriguing late-time properties included persistent P-Cyg absorptions attributable to Fe~II at $< 2000$ \kms, and unusually strong emissions from Ca~II H\&K, Na~I\,D, and O~I 7774 \AA\ (see Figure~\ref{fig:sn2012au}). Persistent P-Cyg absorptions and asymmetries between elements and their ions in the emission line profiles are consistent with expectations of a moderately aspherical and potentially jetted explosion that was most likely initiated by the core collapse of a massive progenitor star. Aside from some differences in the strength and velocity widths of the Fe and Mg emissions, above $\approx 5600$ \AA\ emissions from SN\,2012au are almost indistinguishable from those of hypernovae SN\,1997dq and SN\,1997ef, and the superluminous SN\,2007bi. Like SN 2012au, these objects exhibited slow spectroscopic evolution and slowly declining light curves.

Extensive radio and X-ray observations of SN\,2012au closely follow models of synchrotron emission from a CSM  with wind density profile $\rho \propto r^{-2}$ and mass-loss rate of the progenitor of $\dot M = 3.6 \times 10^{-6} M_{\odot}\,\rm yr^{-1}$ \citep{Kamble14}. Together with the large explosion energy and the large $^{56}$Ni mass, the progenitor of SN 2012au likely had a main-sequence mass $> 20 M_{\odot}$, for which the outer hydrogen envelope had been stripped away but the helium layer still remained. \citet{Milisavljevic13-12au} concluded that a single framework involving the core collapse of a massive progenitor and a subsequent asymmetric explosion could unify subsets of SNe and SLSNe that span $-21 < M_{B} < -17$ mag.

Observations of other extraordinary events have corroborated the notion that connections between some seemingly ordinary Type Ib,c, SLSNe, and SNe associated with long duration gamma-ray bursts (GRBs) may exist. SN 2011kl was associated with the ultra-long-duration gamma-ray burst, GRB 111209A, at a redshift z of 0.677 \citep{Greiner15}. Its light curve  was significantly overluminous compared to other GRB-associated SNe and suggested a link between SN-GRB and SLSNe. Also significant in this regard is nebular-phase spectroscopy of the SLSN-I SN\,2015bn spanning +250--400 days after maximum light. These spectra (among the latest ever obtained for a SLSN) are virtually identical to SN\,2012au and other energetic SNe Ic \citep{Nicholl16}, and supported the \citet{Milisavljevic13-12au} interpretation that relatively narrow O I $\lambda$7774 line may be the signature of a central engine.

\subsection{Calcium-rich transients: {\it iPTF15eqv}}

The class of ``Ca-rich transients'' defined by unusually strong calcium line emissions that develop in optical spectra months after explosion has garnered considerable attention in the last decade. They were first reported by \citet{Filippenko03}. Later,  thorough investigations were provided by \citet{Perets10} of SN\,2005E, and \citet{Kasliwal12} who discovered several examples in PTF survey data.

Ca-rich transients have distinct He lines in their spectra during the first couple months of evolution and fall within the supernova Type Ib classification. However, Ca-rich transients are less luminous than normal SNe Ib and reach nebular stages much more rapidly (within two months post-explosion). Perhaps most intriguing is that Ca-rich transients are often located at large projected distances (as high as 150 kpc) from the centers of their host galaxies, implying that progenitors have traveled significant distances before exploding \citep{Lyman14,Foley15}.  Many occur in early-type galaxies lacking obvious massive star populations \citep{Kasliwal12}, and examinations of explosion sites with deep, high resolution images have thus far failed to uncover any sign of in situ star formation \citep{Lyman13,Lyman14,Lyman16,Lunnan17}. 

\begin{figure}[!htp]
\centering
\includegraphics[width=0.49\linewidth]{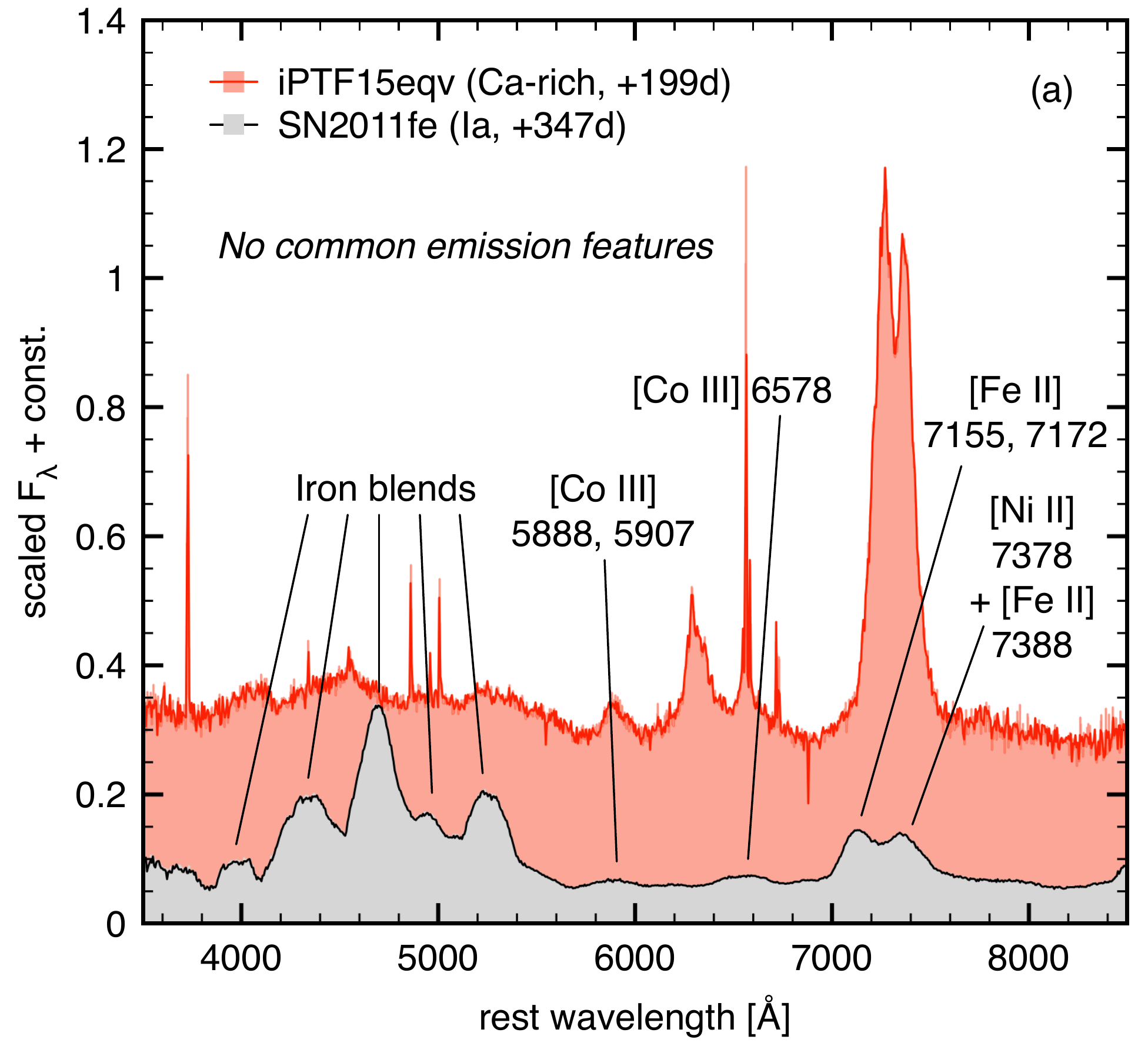}
\includegraphics[width=0.49\linewidth]{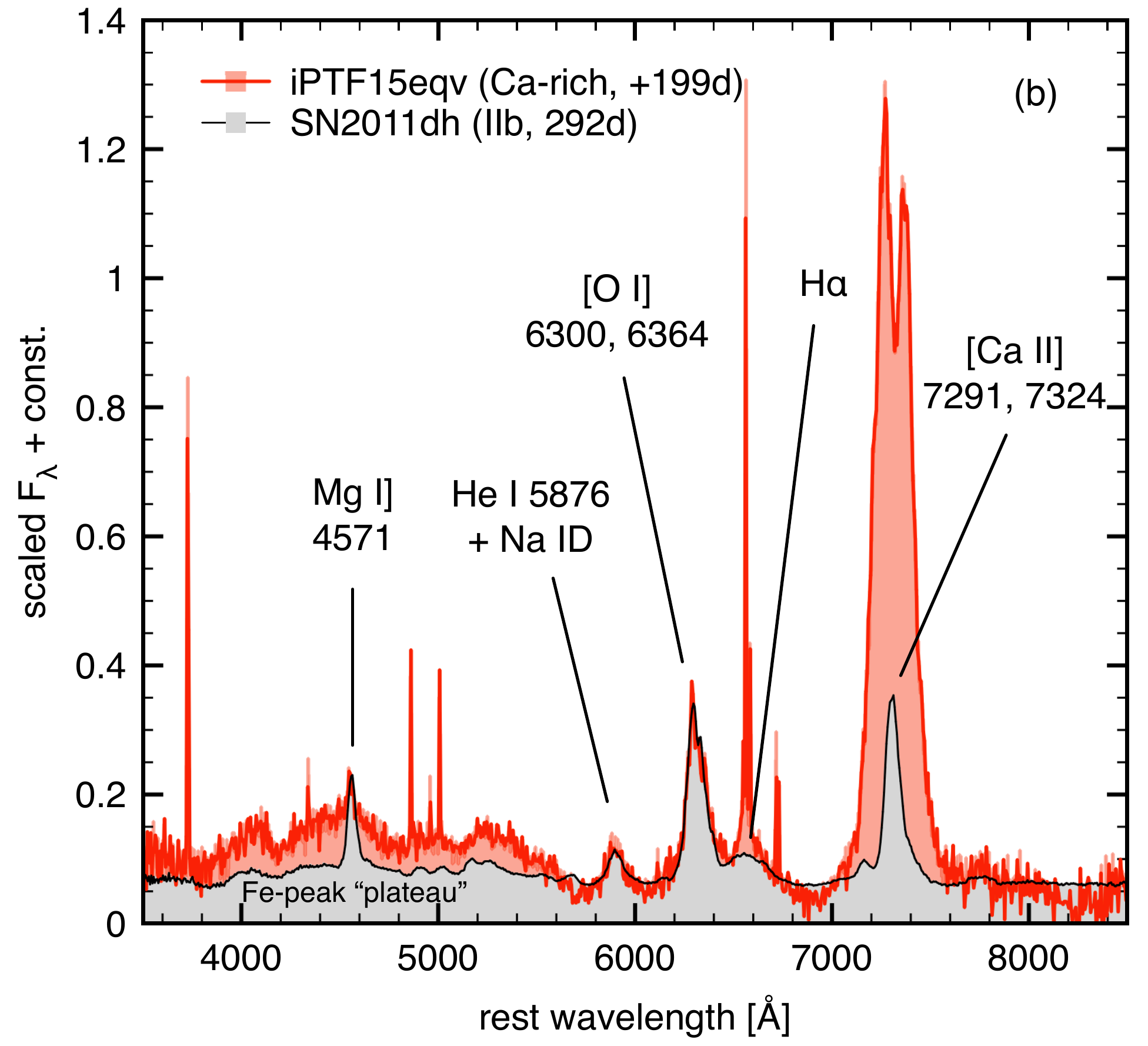}\\
\includegraphics[width=0.49\linewidth]{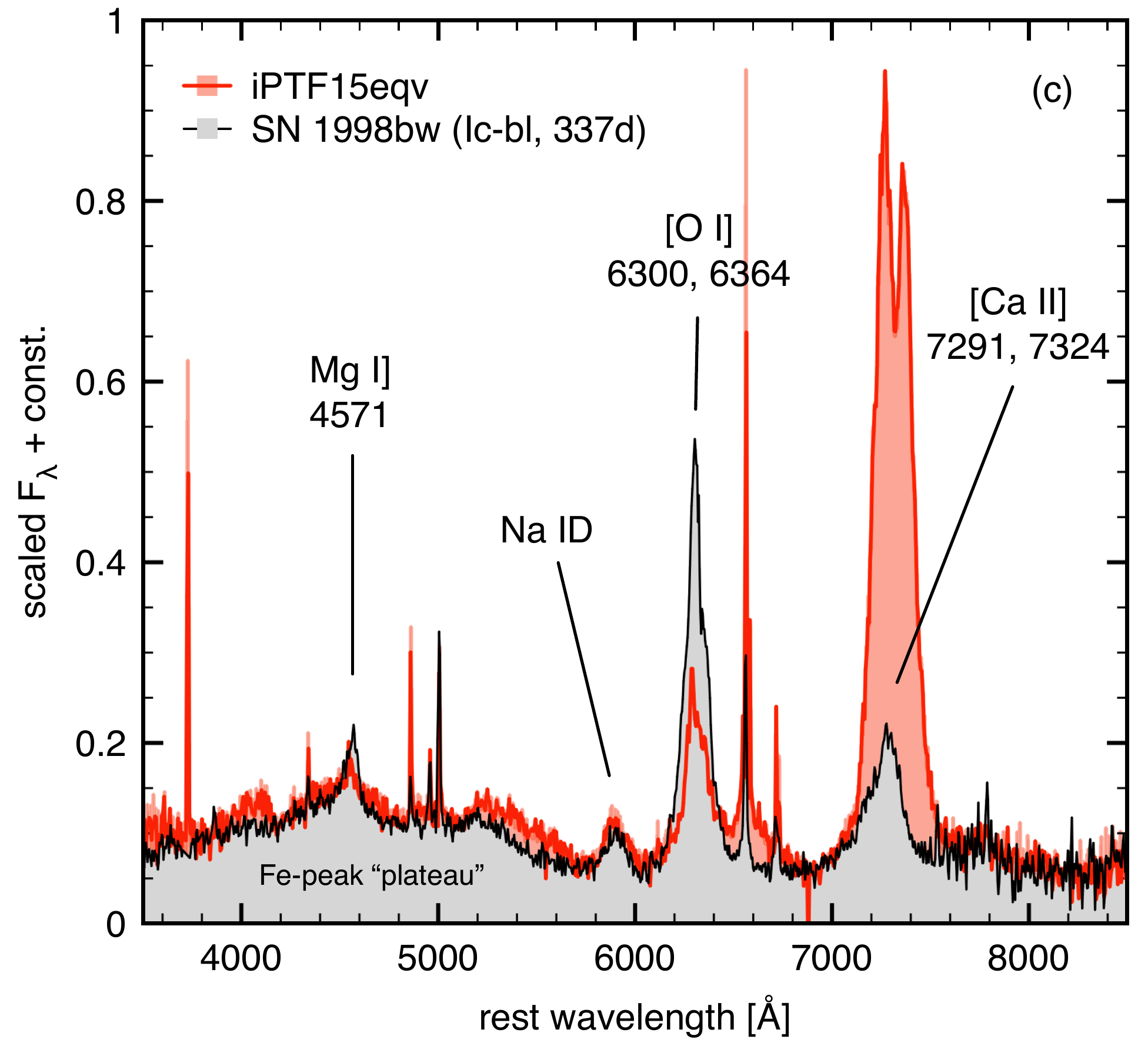}
\includegraphics[width=0.49\linewidth]{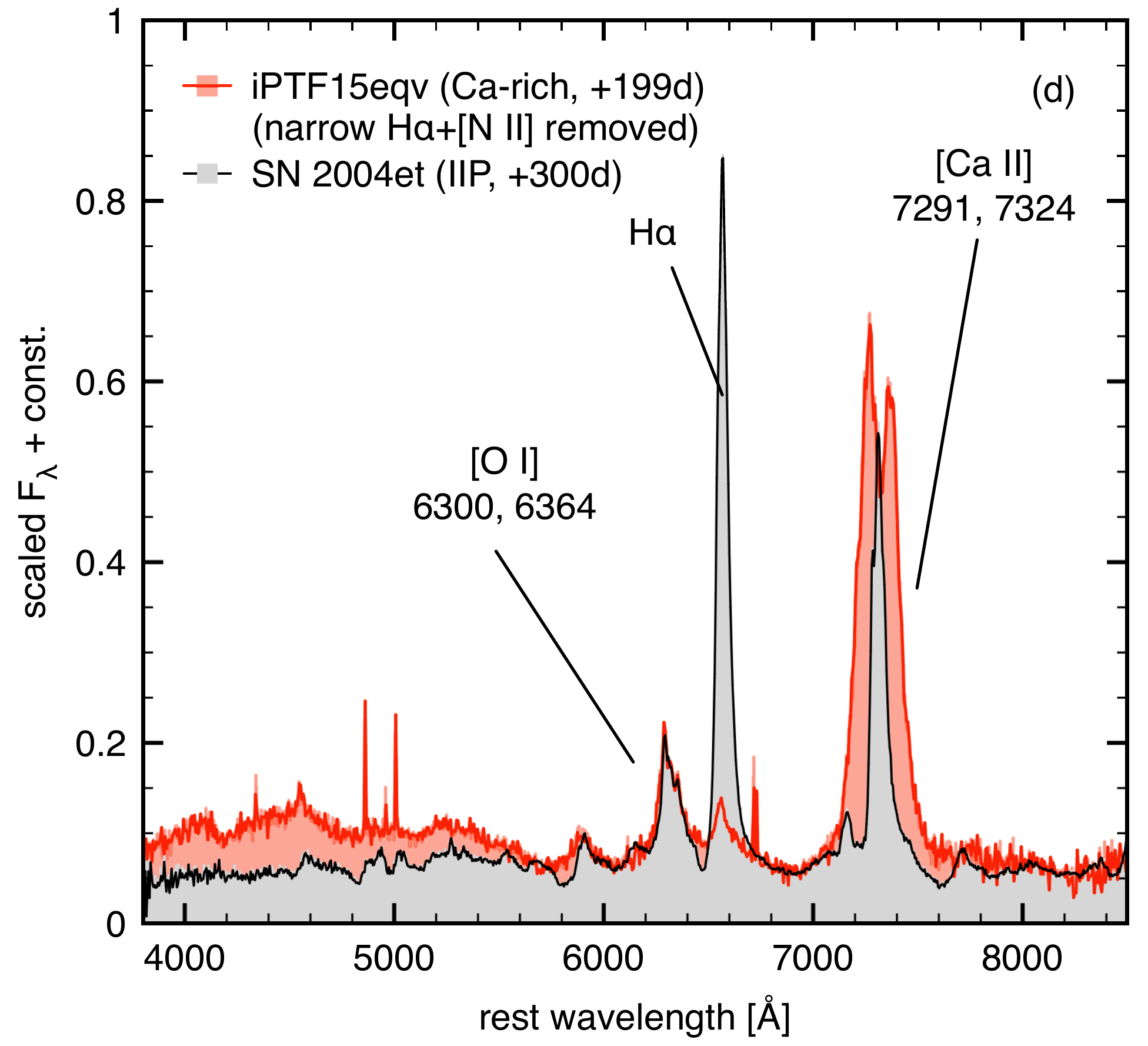}

\caption{Spectral fingerprinting iPTF15eqv. Spectra have been arbitrarily scaled to compare and contrast the relative strengths of specific emission features between nebular epoch spectra of iPTF15eqv and other supernovae (both thermonuclear and core collapse). The spectral features of iPTF15eqv best match those of a core- collapse explosion and have little similarity to those observed in thermonuclear WD explosions. Data were originally published in \citet{Mazzali15}, Ergon et al. (2015), Patat et al. (2001), and Sahu et al. (2006). Adapted from \citet{Milisavljevic17}.}

\label{fig:iPTF15eqv}
\end{figure}

\begin{figure}[!htp]
\centering
\includegraphics[width=0.85\linewidth]{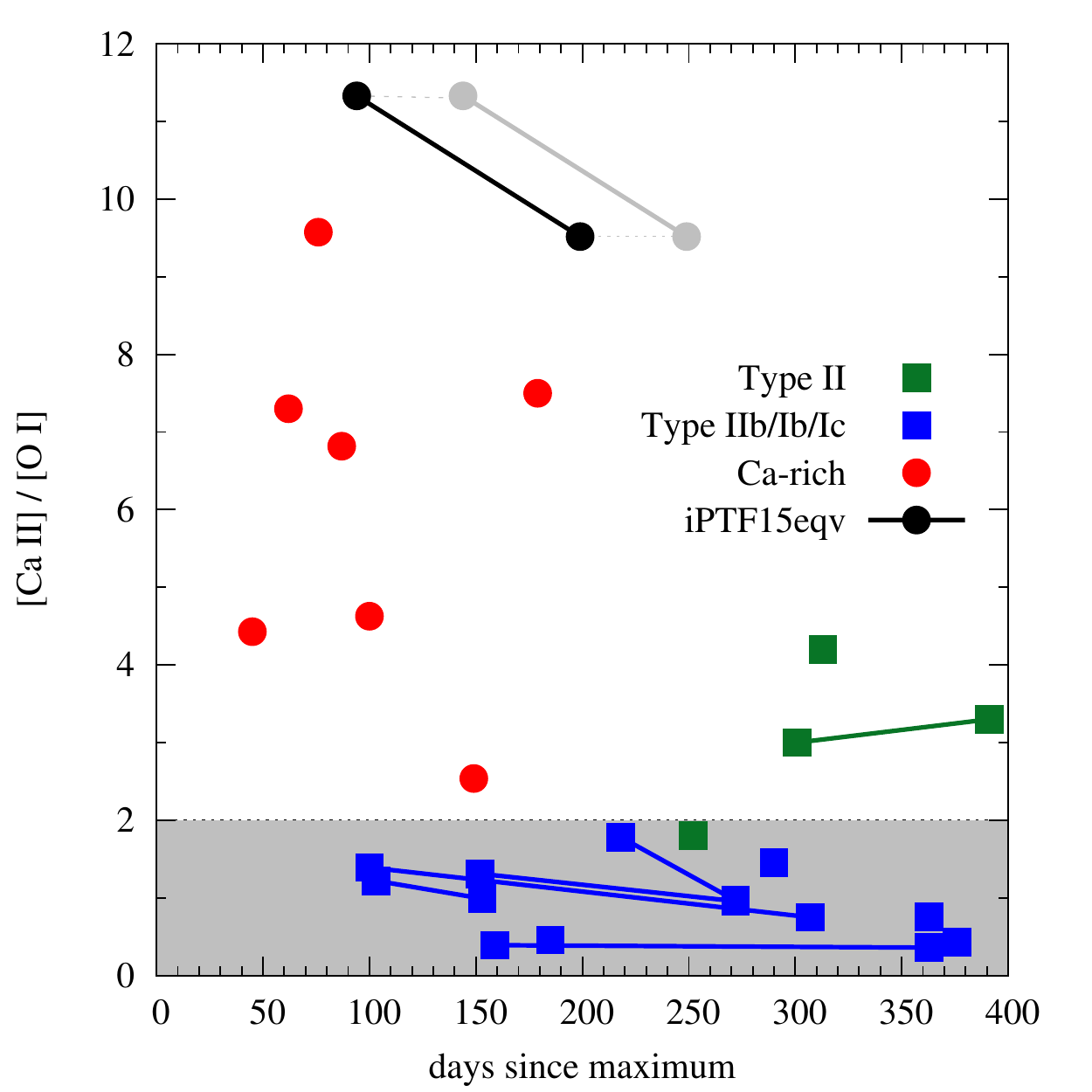}

\caption{Emission line ratio of [Ca II] $\lambda\lambda$7291, 7324 to [O I] $\lambda\lambda$6300, 6364 for Type II, Type IIb/Ib/Ic, and Ca-rich transients. At all epochs where both lines are visible and the conditions can be reasonably assessed to be nebular, all Ca-rich transients have [Ca II]/[O I]$>$2, and all Type Ib/c are $< 2$. Solid lines connect different epochs of the same object. iPTF15eqv stands out with the highest [Ca~II]/[O~I] ratio of $\approx10$. The gray silhouette of iPTF15eqv connected by dashed lines reflects the uncertainty in the explosion date. Adapted from \citet{Milisavljevic17}.}

\label{fig:ca-o-ratio}
\end{figure}

The progenitor systems of Ca-rich transients have been a source of contention.  A variety of spectroscopic properties of Ca-rich transients, including the Type Ib classification, are most naturally understood as originating from explosions of massive star progenitors \citep{Kawabata10}. However, the prevailing view is that Ca-rich transients are somehow related to WD progenitor systems. The relatively large delay-time distribution required to travel large distances favors an older white dwarf (WD) population that also contributes to SNe Ia. Many explosion channels have been proposed involving helium detonations occurring on a helium-accreting WD \citep{SB09,Perets10,WK11}.  Attempts to understand Ca-rich transients using models involving the tidal detonation of a low-mass WD by a neutron star or intermediate-/stellar-mass black hole have also been suggested \citep{Rosswog08,MacLeod14,Sell15}. 

The discovery of iPTF15eqv -- a ``Peculiar Ca-rich Transient'' -- complicated the WD progenitor paradigm and corroborated suspicions that the observational classification may encompass multiple progenitor channels \citep{Milisavljevic17}. The close proximity and relatively slow decline rate of iPTF15eqv enabled spectroscopic observations of high signal-to-noise ratio at epochs $>200$ days after explosion, which is among the latest epochs ever observed for a Ca-rich transient.   The transient was originally discovered by K.\ Itagaki, and then independently discovered and classified by the iPTF survey \citep{Cao15}. Initial spectra led \citet{Cao15} to classify iPTF15eqv as an SN IIb/Ib in the nebular phase. However, a multi-wavelength follow-up campaign supported by radio observations with the  VLA and X-ray observations with {\it Chandra} showed a combination of properties that bridge those observed in Ca-rich transients and SNe Ib/c \citep{Milisavljevic17}.  

Perhaps most revealing about iPTF15eqv is its conspicuous Type Ib/c and Type II spectroscopic signatures in late-time optical and NIR data (Figure~\ref{fig:iPTF15eqv}).  iPTF15eqv exhibits [O~I], [Ca~II], a blend of He I+Na I, and a ``plateau? of emission spanning 3800-5600 \AA\ associated with Fe-peak elements. To varying degrees these are all observed in Type IIb SN 2011dh \citep{Ergon15}, the GRB-SN SN 1998bw, and the Type IIP SN 2004et \citep{Sahu06}. A broad emission feature redward of [O I] that is likely H$\alpha$  but may include contribution from [N~II] and/or Ca~I] $\lambda$6572, is also shared in some cases. In sharp contrast, iPTF15eqv does not share any conspicuous features with normal Type Ia that are dominated by blended forbidden and permitted Fe-peak lines between 4000-5500 \AA\ and 7000-7600 \AA. 

On the one hand, the late-time spectra of iPTF15eqv are dominated by strong calcium emission lines, which is a defining characteristic of Ca-rich transients. Its [CaII]/[OI] $\approx 10$ emission line ratio is among the highest encountered among Ca-rich events (Figure~\ref{fig:ca-o-ratio}). On the other hand, iPTF15eqv differs from other Ca-rich transients in terms of its light curve evolution. iPTF15eqv was a slower evolving and potentially much more luminous object than Ca-rich transients (estimated to potentially be up to $\sim 2$ mag brighter than other examples; \citealt{Kawahara18}) and exhibited a light curve that resembles SNe Ib/c.

Analysis of the chemical abundances associated with observed line emission from iPTF15eqv is consistent with the supernova explosion of a $< 10 M_{\odot}$ star that was stripped of its H-rich envelope via binary interaction. This result challenges the notion that spectroscopically classified Ca-rich transients only originate from WD progenitor systems.   \citet{Milisavljevic17} conclude that the relatively long delay-time distribution and distinguishable abundance patterns of electron capture supernovae from binary systems make them an attractive progenitor system for at least some Ca-rich transients.

\begin{figure}[!htp]
\centering
\includegraphics[width=0.49\linewidth]{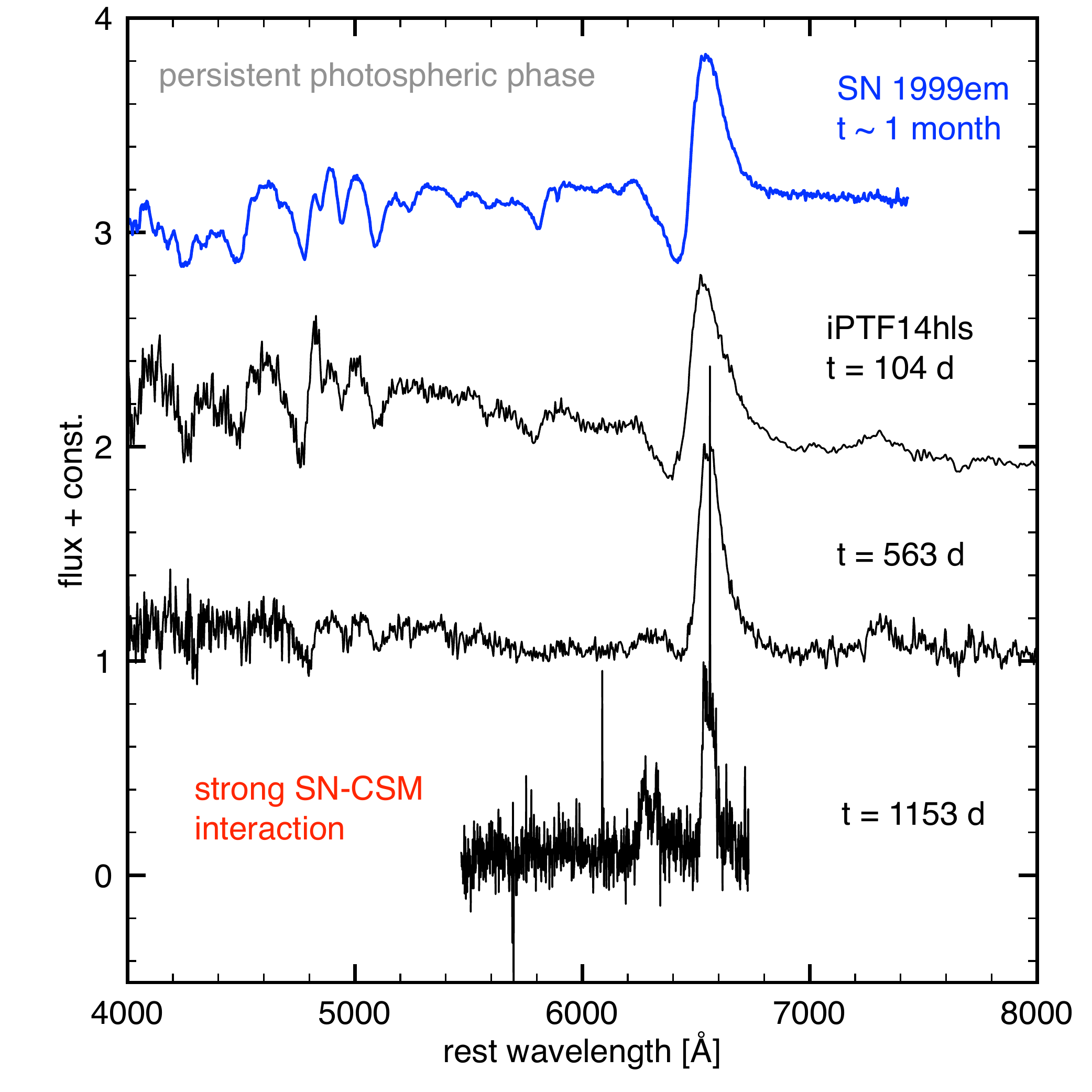}
\includegraphics[width=0.49\linewidth]{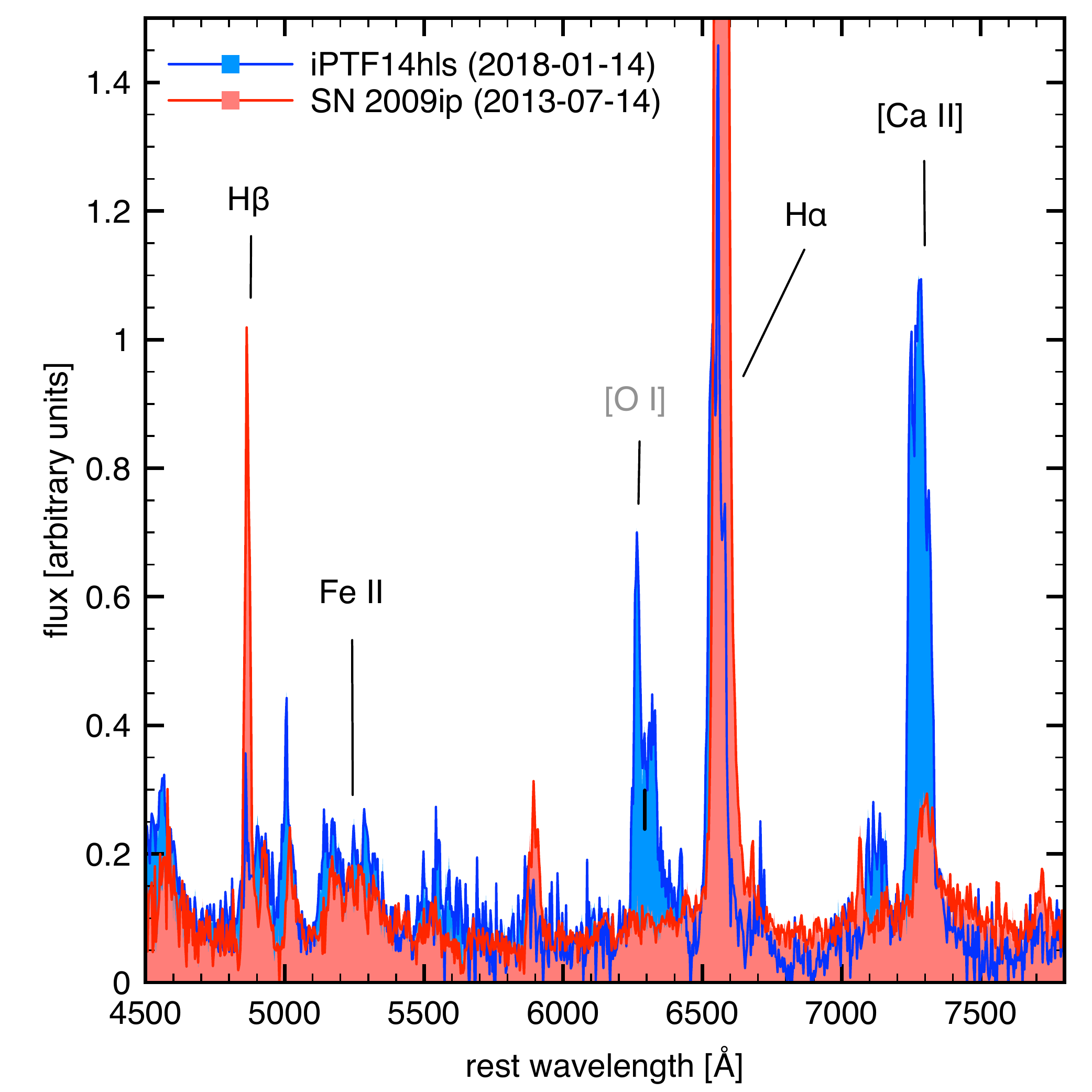}

\caption{{\it Left}: Spectral sequence of iPTF14hls. iPTF14hls exhibited spectral features identical to normal Type II-P explosions (shown here is SN\,1999em for comparison), yet persisted nearly unchanged for $\sim 600$ days while the light-curve experienced multiple re-brightenings. Data are from \citet{Arcavi17} and downloaded from the Weizmann Interactive Supernova data Repository \citep{Yaron12}.  \citet{Andrews17} published a spectrum obtained on day 1153 showing dramatic transformation consistent with strong SN-CSM interaction. {\it Right}: Keck optical spectrum of iPTF14hls obtained 1210 days after discovery (Terreran et al., in preparation) compared to the strongly interacting Type IIn SN\,2009ip \citep{Margutti14}. This spectrum, which spans a wider wavelength window compared to the \citet{Andrews17} data and shows a combination of broad emission lines associated with shocked ejecta and narrow emission lines associated with nearby photoionized material, is consistent with their interpretation that the dramatic change is the consequence of SN-CSM interaction. Notably, the broad [O I] 6300, 6364 emission line feature, not observed (and possibly obscured) in SN\,2009ip, signifies considerable metal-rich O-rich ejecta in iPTF14hls that is interacting with a disk-like environment. }

\label{fig:14hls}
\end{figure}

\subsection{Unusual Longevity: {\it iPTF14hls}}

iPTF14hls \citep{Arcavi17} commanded recent attention for exhibiting both extremely normal and extremely peculiar characteristics simultaneously. iPTF14hls showed spectra that are identical to normal Type II-P explosions, yet persisted nearly unchanged for $\sim 600$ days while the light-curve experienced multiple re-brigthenings (Figure~\ref{fig:14hls}).  The absorption lines showed negligible decreases in velocity throughout the five peaks of the luminous light curve. A pre-explosion outburst in 1954 was also reported. 

The spectral and temporal evolution of iPTF14hls is unprecedented. Possible frameworks to understand its nature include magnetar central engines and pulsational pair-instability supernovae \citep{Dessart18,Woosley18}, or common envelope jets \citep{Soker18}.  High-energy $\gamma$-ray emission may have been detected \citep{Yuan17}, which suggests very fast particle acceleration by the supernova explosion. 

A significant breakthrough in constraining the nature of iPFT14hls came with an abrupt change in its spectral emissions observed in data obtained by \citet{Andrews17} three years after discovery. A double-peaked intermediate-width H$\alpha$ line indicative of expansion speeds around 1000 \kms\ was observed. A similar profile was also observed in the [O~I] 6300, 6364 lines. \citet{Andrews17} interpreted this as clear evidence of interaction between the SN and dense CSM having a disc-like geometry, which is an important clue for understanding the nature of the explosion. They conclude that interaction with variations in the density structure of the CSM may be adequate to explain the peculiar evolution of iPTF14hls.  A later epoch Keck spectrum obtained by Terreran et al.\ (in preparation) shows a combination of narrow and broad emission lines seen in strongly interacting Type IIn supernovae consistent with the \citet{Andrews17} SN-CSM interpretation (Figure~\ref{fig:14hls}). Further monitoring across wavelengths will hopefully continue to shed additional light on the nature of this exciting transient. 

\section{Special case: ASASSN-15lh}

\begin{figure}[!htp]
\centering
\includegraphics[scale=0.35]{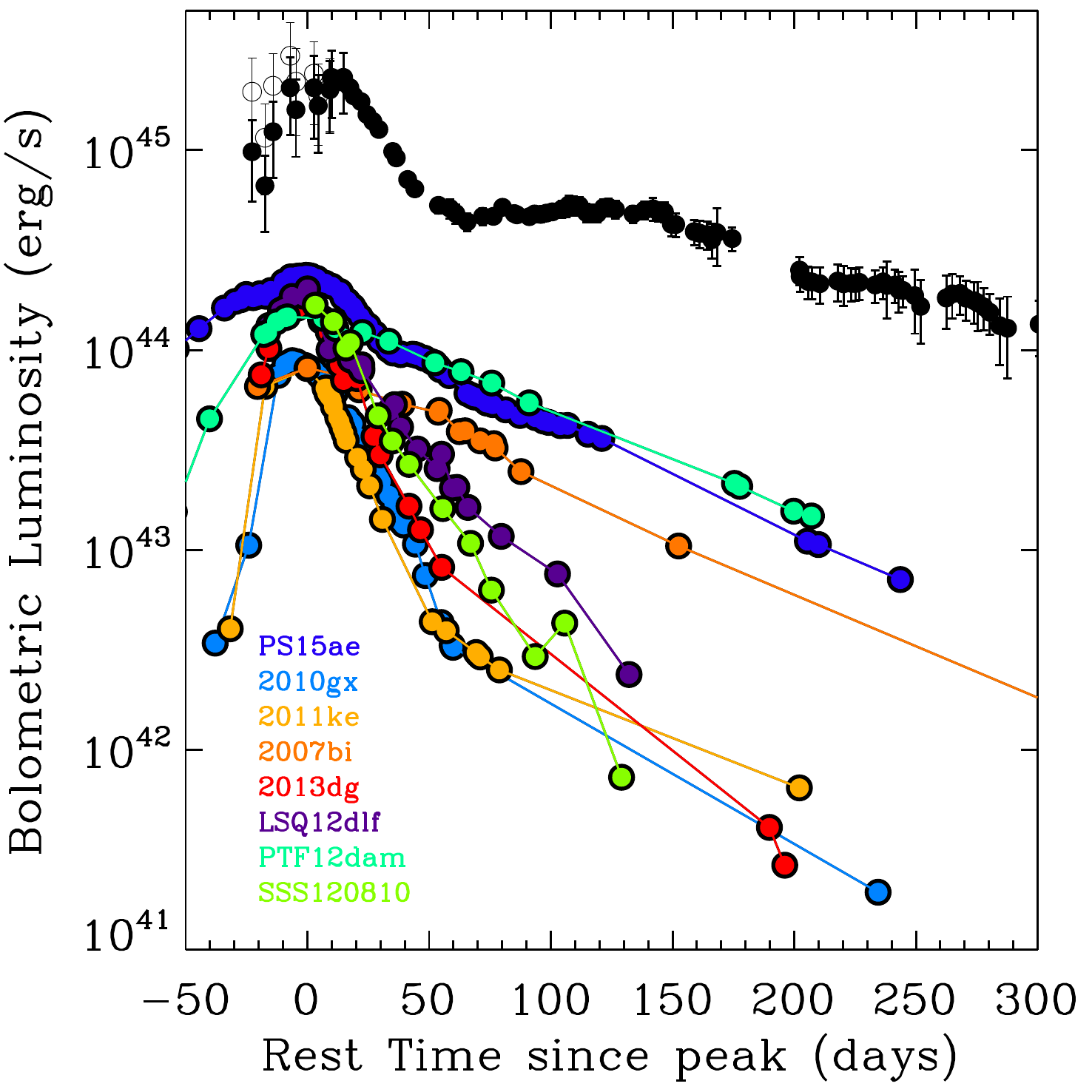}
\includegraphics[scale=0.3]{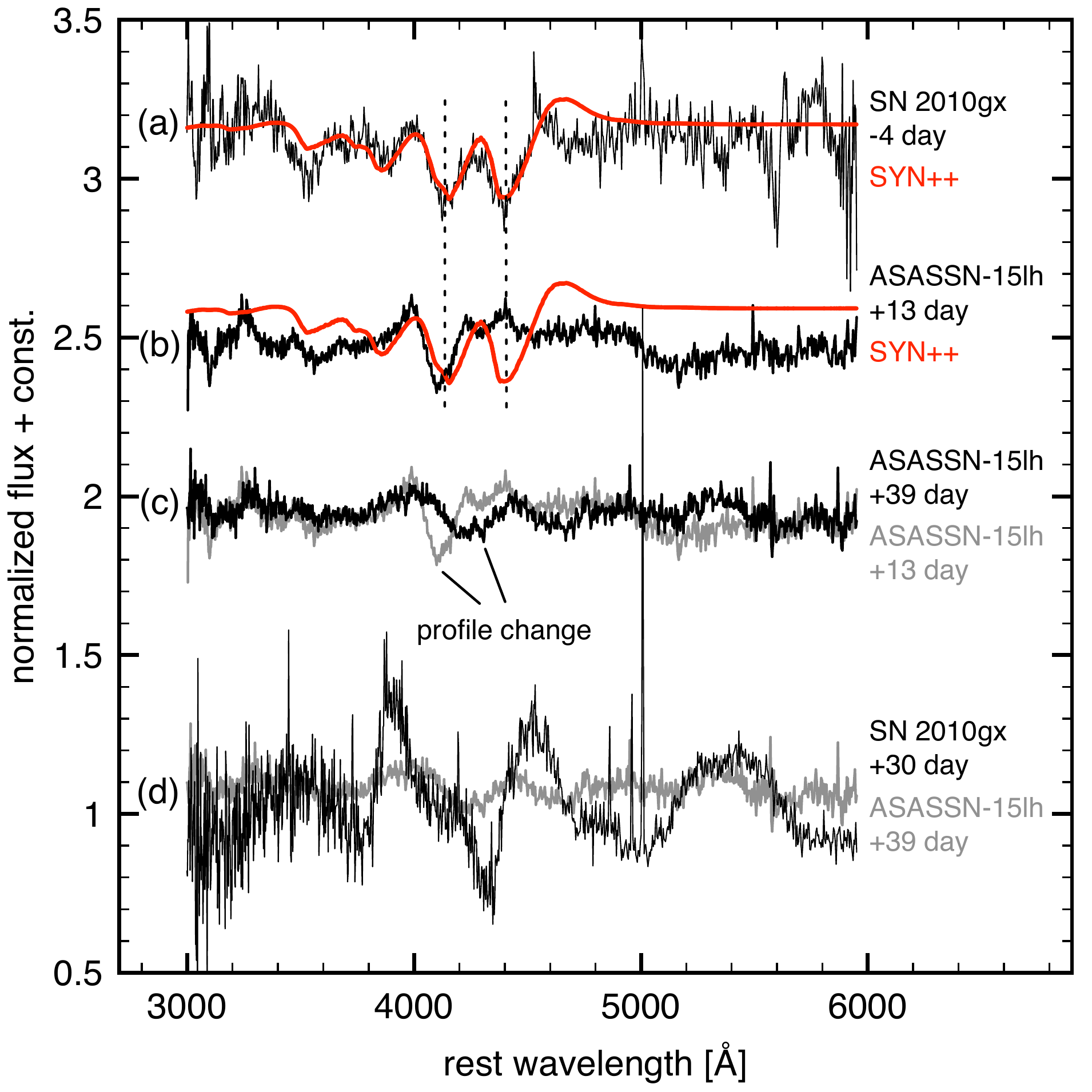}
\caption{\emph{Left panel:} Bolometric luminosity emission from ASASSN-15lh (black dots from \citealt{Dong16,Margutti17}) compared to a sample of H-stripped SLSNe (from \citealt{Inserra13,Nicholl16}). ASASSN-15lh is $>10$ times more luminous then the most luminous SLSNe known at peak and shows an unusually very mild decay at later times. \emph{Right panel:} early-times spectral evolution of ASASSN-15lh in context with SLSNe (here we use the SLSN SN2010gx, which shows the typical features of the class). The O II ion ``W-shaped" feature is typical of SLSNe (vertical dashed lines), and here we show how using the simple assumptions of SYN++ and a photospheric velocity of 19,000 km~s$^{-1}$, the -4 day spectrum of the SLSN 2010gx can be reproduced. By contrast, we cannot reproduce +13 day spectrum of ASASSN-15lh. Even more importantly, ASASSN-15lh progressively  evolves towards a featureless spectrum with time, which is the opposite evolution of SNe. Adapted from \cite{Margutti17}.}
\label{fig:15lh-lc}
\end{figure}

The transient ASASSN-15lh is an excellent example of how some phenomena fail to fit naturally into any known classification. Indeed, the properties of ASASSN-15lh are so peculiar that, at the time of writing, it is not even clear if ASASSN-15lh is connected with a stellar explosion and might be the first (and only member so far) of its own class.

ASASSN-15lh  was discovered \citep{Dong16} by the All- Sky Automated Survey for Supernovae (ASAS-SN) at z=0.2326  ($d=1171$ Mpc for standard \emph{Planck} cosmology). The location of ASASSN-15ls is astrometrically consistent with he nucleus of its massive early type host galaxy (\citealt{Dong16}). The transient exhibited an extremely large peak luminosity $L_{\rm pk} \sim 2 \times 10^{45}$\,erg\,s$^{-1}$ and a blue, almost featureless spectrum with no apparent sign of H or He, leading \citet{Dong16} and \citet{Godoy-Rivera17} to suggest it to be the most luminous SLSN ever detected (Fig. \ref{fig:15lh-lc}). However, the spectral and temporal evolution of the transient, as well as its host galaxy, are unprecedented among SLSNe and other explanations have been proposed, including a tidal disruption event (TDE) by a super-massive BH \citep{leloudas16}. 

The observed properties of ASASSN-15lh challenge any classification scheme and can be summarized as follows:
\begin{itemize}
\item The peak luminosity of ASASSN-15lh and its total radiated energy ($E_{\rm{rad}}\sim2\times 10^{52}\,\rm{erg}$) are extreme even among SLSNe, and require sources of energy that are different from the standard radioactive decay of $^{56}$Ni that powers normal H-stripped SNe in the local Universe (\citealt{Dong16,Chatzopoulos16,Kozyreva16,vanPutten16}). 
\item ASASSN-15lh experienced a UV rebrightening beginning at $t \sim 90$-d (observer frame) after the primary peak and was followed by a $\sim 120$-d long plateau in the bolometric luminosity (Fig. \ref{fig:15lh-lc}), before starting to fade again. Throughout its initial decline, subsequent rebrightening and renewed decline, the spectra did not show evidence of interactions between the ejecta and CSM such as narrow emission lines. There are hints of weak H$\alpha$ emission at late-times, but \citet{Margutti17} have shown that it is narrow line emission consistent with star formation in the host nucleus. 
\item Differently from SNe, ASASSN-15lh evolved towards a progressively featureless spectrum (Fig. \ref{fig:15lh-lc}).
\item The host galaxy of ASASSN-15lh is old, massive $M_{\ast}\sim 2\times 10^{11}\,\rm{M_{\odot}}$ and with limited star formation \citep{Dong16}. These properties are very different from galaxies that host core-collapse SNe and SLSNe, which are typically younger star forming galaxies with significantly lower stellar mass (\citealt{Lunnan14,Lunnan15,Perley16,leloudas16}). In the context of TDEs, the host galaxy properties are also unprecedented, and ASASSN-15lh would be the most luminous TDE ever observed, associated with a SMBH with mass $M_{\bullet}\sim10^{8.6}\,\rm{M_{\odot}}$ significantly larger than the average TDE SMBHs (e.g. \citealt{Komossa15}). The large SMBH mass motivated speculation of a fast rotating SMBH, in order to disrupt the star outside the innermost stable stellar orbit \citep{leloudas16,Margutti17}. \citet{Coughlin17} argue that its features, including its anomalous rebrightening at $\sim 100$ days after detection, are consistent with the tidal disruption of a star by a supermassive black hole in a binary system.
\end{itemize}

Debate has continued on whether ASASSN-15lh is an H-poor superluminous supernova or a tidal disruption event. It is clear that the properties of ASASSN-15lh and of its host galaxy are unprecedented in both scenarios. It is possible that a clue to the real nature of ASASSN-15lh has already been provided by its emission outside the optical range. A faint X-ray source has been detected at the location consistent with ASASSN-15lh \citep{Margutti17}. The very soft spectrum of the source and its persistent emission are not consistent with SNe, and suggest instead an origin connected with the nucleus of the host galaxy. If ASASSN-15lh is a TDE, the expectation is fading of the X-ray source in the next few years. 

Future multi-wavelength follow-up of ASASSN-15lh, and a larger sample of similar events, are the only way to advance our understanding of the physics of this extremely peculiar transient.

\section{Concluding Remarks: what will be ``peculiar'' in the future?}
\label{sec:conclusions}

In this chapter we have reviewed several examples of peculiar supernovae in an effort to understand what sets them apart from normal events. A general pattern emerged that classes of supernovae that were originally interpreted to have unusual properties are later seen as logical branches of phenomena that are already well understood. In some cases, the nature of a peculiar transient can remain debated, and we discussed the special case of ASASSN-15lh as an example of this contentious variety. The shift in interpretation from  extraordinary to mundane has often been brought about by improvements in telescope and detector technology, and by observations that push into new phase spaces of luminosity, time scale, and wavelength. 

{\it What will be identified as peculiar in the future?} Observations of supernovae in the first hours to days following explosion have been limited thus far but provide enticing glimpses at the wealth of information that can be extracted at extremely early phases \citep{Nugent11,Milisavljevic13-11ei,Gal-Yam14,Nicholl15,Garnavich16}. The Pan-STARRS, PTF, and $D < 40$ Mpc (DLT40) surveys have made commendable progress in this regard \citep{Drout14,Yaron17,Tartaglia18}, and the upcoming surveys Zwicky Transient Facility\footnote{\url{https://www.ptf.caltech.edu/ztf}} and Large Synoptic Survey Telescope\footnote{\url{https://www.lsst.org/}} (LSST) will allow even greater exploration of these very early phases. Complementary to these advances will be explorations at very late epochs. This will be a particular strength of LSST, which will provide a deep, all-sky survey sampled in a consistent rapid cadence, and next-generation telescopes such as the Giant Magellan Telescope\footnote{\url{https://www.gmto.org/}}, the Thirty Meter Telescope\footnote{\url{https://www.tmt.org/}}, and the European Extremely Large Telescope\footnote{\url{https://www.eso.org/sci/facilities/eelt/}} that will revolutionize late-time investigations of supernovae both in the volume of space that can be sampled (and hence the number of objects to study) and the quality of data to be obtained. Particularly exciting will be the ability to carefully monitor the evolution of emission line widths in spectra, which can distinguish between sources of energy in extreme supernovae \citep{Milisavljevic12}. As these and other facilities, now operating in the multi-messenger era \citep{Abbott17}, move into uncharted frontiers of transient phase space, they are destined to continue finding new examples of supernovae that defy expectations and shape our understanding of the terminal stages of stellar evolution.

\begin{acknowledgements}

D.~M.\ \& R.~M.\ thank ISSI organizers for their kind invitation to the Supernova Workshop in Bern. D.~M.\ acknowledges support from NASA through grant number GO-14202 from the Space Telescope Science Institute, which is operated by AURA, Inc., under NASA contract NAS 5-26555.  R.~M.\  acknowledges partial support to her group from NASA through NuSTAR grants NNX17AI13G and NNX17AG80G. Partial support for this work was provided by the National Aeronautics and Space Administration through Chandra Award Number GO5-16064A and GO6-17054A issued by the Chandra X-ray Center, which is operated by the Smithsonian Astrophysical Observatory for and on behalf of the National Aeronautics Space Administration under contract NAS8-03060. We especially thank Roger Chevalier for his help and patience preparing the manuscript. We thank J.\ Andrews and N.\ Smith for sharing their iPTF14hls spectrum before publication. We also thank G.\ Terreran for sharing Keck observations of iPTF14hls. This work is based in part on observations from the Low Resolution Imaging Spectrometer at the Keck-1 telescope. We are grateful to the staff at the Keck Observatory for their
assistance, and we extend special thanks to those of Hawaiian
ancestry on whose sacred mountain we are privileged
to be guests. The W. M. Keck Observatory is operated as a
scientific partnership among the California Institute of Technology,
the University of California, and NASA; it was made
possible by the generous financial support of the W. M. Keck
Foundation.

 \end{acknowledgements}


%
%


\end{document}